%% file: EHF_Champions_League_simulation_v3.tex
\documentclass[12pt]{article}
\usepackage[latin1]{inputenc}
\usepackage[british]{babel}
\usepackage{cmap}
\usepackage{lmodern}

\usepackage{amssymb, amsmath, amsthm}
\usepackage[a4paper,top=25mm,bottom=25mm,left=25mm,right=25mm]{geometry}
\usepackage{etex}
\usepackage{ragged2e}

\usepackage{authblk} 
\usepackage{pifont}
\usepackage{graphicx}
\usepackage[usenames,dvipsnames,svgnames,table]{xcolor}
\usepackage[figuresright]{rotating}
\usepackage{xtab} 
\usepackage{longtable} 
\usepackage{multirow}
\usepackage{footnote}
\usepackage[stable]{footmisc}
\usepackage{chngpage} 
\usepackage{pdflscape} 
\usepackage[nottoc,notlot,notlof]{tocbibind} 

\usepackage{pgfplots}
\pgfplotsset{every tick label/.append style={font=\small}}
\pgfplotsset{compat=1.14}
\usepackage{setspace}

\makesavenoteenv{tabular}
\usepackage{tabularx}
\usepackage{booktabs}
\usepackage{threeparttable}
\usepackage[referable]{threeparttablex} 
\newcolumntype{R}{>{\raggedleft\arraybackslash}X}
\newcolumntype{L}{>{\raggedright\arraybackslash}X}
\newcolumntype{C}{>{\centering\arraybackslash}X}
\newcolumntype{A}{>{\columncolor{gray!25}}C}
\newcolumntype{a}{>{\columncolor{gray!25}}c}

\newlength{\tablen}

\usepackage{dcolumn} 
\newcolumntype{.}{D{.}{.}{-1}}

\usepackage{tikz}
\usetikzlibrary{arrows, calc, matrix, patterns, positioning, trees}
\usepackage[semicolon]{natbib}
\usepackage[hyphens]{url}
\usepackage{hyperref} 
\hypersetup{
  colorlinks   = true,    
  urlcolor     = blue,    
  linkcolor    = blue,    
  citecolor    = ForestGreen      
}
\usepackage{microtype}
\usepackage[justification=centering]{caption} 

\usepackage[labelformat=simple]{subcaption}

\DeclareCaptionLabelFormat{parenthesis}{(#2)}
\makeatletter
\renewcommand\p@subfigure{\arabic{figure}.}
\makeatother

\DeclareCaptionLabelFormat{parenthesis}{(#2)}
\makeatletter
\renewcommand\p@subtable{\arabic{table}.}
\makeatother

\usepackage{enumitem}

\setlist[itemize]{leftmargin=2.5\parindent}
\setlist[enumerate]{leftmargin=2.5\parindent}

\pgfplotsset{every tick label/.append style={font=\footnotesize}}

\newcommand\Pair[3]{%
  \begin{tabular}{|>{\centering\arraybackslash}m{0.75cm}|>{\centering\arraybackslash}m{1.5cm}|}
  \hline
  \multirow{2}{*}{#1} & #2 \\ \cline{2-2}
   & #3 \\
  \hline
  \end{tabular}%
}
\newcommand\WidePair[3]{%
  \begin{tabular}{|>{\centering\arraybackslash}m{0.75cm}|>{\centering\arraybackslash}m{2cm}|}
  \hline
  \multirow{2}{*}{#1} & #2 \\ \cline{2-2}
   & #3 \\
  \hline
  \end{tabular}%
}

\theoremstyle{plain}

\theoremstyle{definition}

\theoremstyle{remark}

\def\keywords{\vspace{.5em} 
{\noindent \textit{Keywords}: }}

\def\JEL{\vspace{.5em} 
{\noindent \textbf{\emph{JEL} classification number}: }}

\def\AMS{\vspace{.5em} 
{\noindent \textbf{\emph{MSC} class}: }}

\author{\href{https://sites.google.com/view/laszlocsato}{L\'aszl\'o Csat\'o}\thanks{~E-mail: \emph{laszlo.csato@sztaki.mta.hu}} }
\affil{Institute for Computer Science and Control, Hungarian Academy of Sciences (MTA SZTAKI) \\
Laboratory on Engineering and Management Intelligence, Research Group of Operations Research and Decision Systems}
\affil{Corvinus University of Budapest (BCE) \\
Department of Operations Research and Actuarial Sciences}
\affil{Budapest, Hungary}
\title{Optimal tournament design: lessons from \\ the men's handball Champions League}
\date{\today}

\def\Dedication{
{\noindent
$\mathfrak{Der}$ $\mathfrak{Beweis}$, $\mathfrak{den}$ $\mathfrak{wir}$ $\mathfrak{fordern}$, $\mathfrak{ist}$ $\mathfrak{\ddot{u}berall}$ $\mathfrak{n\ddot{o}tig}$, $\mathfrak{wo}$ $\mathfrak{der}$ $\mathfrak{Vorzug}$ $\mathfrak{des}$ $\mathfrak{vorgeschlagenen}$ $\mathfrak{Mittels}$ $\mathfrak{nicht}$ $\mathfrak{so}$ $\mathfrak{evident}$ $\mathfrak{ist}$, $\mathfrak{da\ss}$ $\mathfrak{er}$ $\mathfrak{keinen}$ $\mathfrak{Zweifel}$ $\mathfrak{zul\ddot{a}\ss t}$, $\mathfrak{und}$ $\mathfrak{er}$ $\mathfrak{besteht}$ $\mathfrak{darin}$, $\mathfrak{da\ss}$ $\mathfrak{jedes}$ $\mathfrak{der}$ $\mathfrak{beiden}$ $\mathfrak{Mittel}$ $\mathfrak{seiner}$ $\mathfrak{Eigent\ddot{u}mlichkeit}$ $\mathfrak{nach}$ $\mathfrak{untersucht}$ $\mathfrak{und}$ $\mathfrak{mit}$ $\mathfrak{dem}$ $\mathfrak{Zweck}$ $\mathfrak{verglichen}$ $\mathfrak{werde}$.\footnote{~
``\emph{The demonstration we require is always necessary when the superiority of the means propounded is not so evident as to leave no room for doubt, and it consists in the examination of each of the means on its own merits, and then of its comparison with the object desired.}'' (Source: Carl von Clausewitz: \emph{On War}, Book 2, Chapter 5---Criticism. Translated by Colonel James John Graham, London, N. Tr\"ubner, 1873. \url{http://clausewitz.com/readings/OnWar1873/TOC.htm})}
}

\flushright
\noindent (Carl von Clausewitz: \emph{Vom Kriege})

\vspace{1cm} 
\justify }

\begin{document}
\newgeometry{top=15mm,bottom=20mm,left=25mm,right=25mm}

\maketitle
\thispagestyle{empty}
\Dedication

\begin{abstract}
\noindent
Many sports tournaments are organised in a hybrid design consisting of a round-robin group stage followed by a knock-out phase. The traditional seeding regime aims to create balanced groups roughly at the same competition level but may result in several uneven matches when the quality of the teams varies greatly.
Our paper is the first challenging this classical solution through the example of the men's EHF (European Handball Federation) Champions League, the most prestigious men's handball club competition in Europe, which has used unbalanced groups between the 2015/16 and 2019/20 seasons. Its particular design is compared to an alternative format with equally strong groups, as well as to the previous scheme of the EHF Champions League. We find that it is possible to increase the quality of all matches played together with raising the uncertainty of outcome, essentially without sacrificing fairness.
Our results have useful implications for the governing bodies of major sports.


\JEL{C44, C63, Z20}

\AMS{62F07, 68U20}

\end{abstract}

\clearpage
\newgeometry{top=25mm,bottom=25mm,left=25mm,right=25mm}





\section{Introduction} \label{Sec1}

The choice of an appropriate format for sports tournaments poses an important question in economics and operations research because ``\emph{designing an optimal contest is both a matter of significant financial concern for the organisers, participating individuals, and teams, and a matter of consuming personal interest for millions of fans}'' \citep[p.~1137]{Szymanski2003}.
In addition, any sports tournament can be considered a kind of selection mechanism with a number of managerial applications such as recruitment strategies \citep{Ryvkin2010}.

Many sports competitions are organised in a hybrid design consisting of a round-robin group stage followed by a knock-out phase. Examples include the \href{https://en.wikipedia.org/wiki/FIFA_World_Cup}{FIFA World Cup}, the \href{https://en.wikipedia.org/wiki/UEFA_Champions_League}{UEFA Champions League}, the \href{https://en.wikipedia.org/wiki/UEFA_European_Championship}{UEFA European Championship} (all in association football), the \href{https://en.wikipedia.org/wiki/FIBA_Basketball_World_Cup}{FIBA Basketball World Cup}, the \href{https://en.wikipedia.org/wiki/FIVB_Volleyball_Men\%27s_World_Championship}{FIVB Volleyball World Championship}, the \href{https://en.wikipedia.org/wiki/European_Men\%27s_Handball_Championship}{EHF European Handball Championship}, or the \href{https://en.wikipedia.org/wiki/IHF_World_Men\%27s_Handball_Championship}{IHF World Handball Championship}.

Since only the top teams from each group qualify for the next stage, the probability of advancing is strongly influenced by the opponents. Therefore, the allocation of the teams into groups is governed by certain rules, and almost all systems use seeds.
The classical solution is to rank the contestants based on their past performance. Then, in the case of $k$ groups, each group gets one team from the first pot of the best $k$ teams, one team from the second pot consisting of the next $k$ teams, and so on.
Further considerations may play a role, for instance, clubs from the same national association could not be drawn against each other in the UEFA Champions League. Similarly, FIFA strives for creating geographically diverse groups in the World Cup.

The seeding procedure above aims to provide balanced groups roughly at the same competitive level, but may inevitably lead to several uneven matches when the quality of the teams varies greatly, which is the usual case.
This can be against the interest of the administrators because higher contest quality and greater uncertainty of outcome are usually associated with higher attendance \citep{ForrestSimmons2002, BorlandMacDonald2003}. Despite that, while there are some recent results on the draw procedures of the groups \citep{Guyon2015a, LalienaLopez2019, CeaDuranGuajardoSureSiebertZamorano2020}, we do not know any work challenging the traditional seeding regime.

Therefore, the current paper attempts to outline and evaluate an alternative to the well-established format of balanced groups. In particular, we will compare the design applied by the men's handball \href{https://en.wikipedia.org/wiki/EHF_Champions_League}{EHF Champions League} from the 2015/16 season with a traditional variant and the previous format of this tournament. Simulations show that the quality of all matches played can be increased and the uncertainty of outcome can be raised, essentially without sacrificing fairness. The latter is crucial because pre-tournament tanking cannot be excluded if strong teams prefer to be drawn into a weak group of a tournament containing unbalanced groups.
Thus our results have useful implications for the governing bodies of major sports.

The paper has the following structure.
Section~\ref{Sec2} describes the three tournament designs to be analysed.
The simulation experiment and the metrics used for the evaluation of the competition formats are presented in Section~\ref{Sec3}.
Section~\ref{Sec4} demonstrates the results and provides a robustness check.
Finally, Section~\ref{Sec6} summarises our main findings.

\section{An instructive competition design} \label{Sec2}

The \href{https://en.wikipedia.org/wiki/EHF_Champions_League}{EHF Champions League}, organised by the European Handball Federation (EHF) since the 1993/94 season, is the most prestigious men's handball club competition of the continent.
The tournament uses a hybrid design mixing knock-out and round-robin stages.
As home advantage is a well-documented phenomenon in this sport \citep{MeletakosBayios2010, Lago_etal2013}, teams usually play both at home and away against each other, with the exception of the Final Four, which takes place at the \href{https://en.wikipedia.org/wiki/Lanxess_Arena}{Lanxess Arena} in Cologne, Germany from the 2009/10 season onwards.

On 21 March 2014, the 119th meeting of the EHF Executive Committee decided to introduce a new competition format from the \href{https://en.wikipedia.org/wiki/2015\%E2\%80\%9316_EHF_Champions_League}{2015/16 season} \citep{EHF2014c}. The number of competing teams was increased in order to open up the tournament to more nations and new markets across the continent. The reform also guaranteed more top matches between the leading clubs for the sake of making the competition more attractive to spectators, sponsors, and the media. At the same time, it was ensured that all clubs playing in the group phase retain the chance to qualify for the Final Four.

This format of the EHF Champions League starts with $28$ competing teams. They play in four round-robin groups, groups A and B with eight teams each, and groups C and D with six teams each such that:
\begin{itemize}
\item
in groups A and B, the top team directly qualifies for the quarter-finals, the bottom two clubs are eliminated, while the remaining teams advance to the first knock-out phase;
\item
in groups C and D, the bottom four teams drop out of the tournament, while the top two teams in both groups contest a play-off to determine the two teams that advance to the first knock-out phase.
\end{itemize}

Consequently, the first knock-out phase involves $12$ teams, five from group A, five from group B, and two from groups C and D. The six winners qualify for the quarter-finals, where they join the group winners of groups A and B. The winners of the quarter-finals participate in the Final Four.

Figure~\ref{Fig_A1} presents the tournament format in detail until the teams of the Final Four are selected. The design remains deterministic after the groups are drawn. For the semi-finals in the Final Four, there is a new draw with all teams being in the same pot.

This competition format will be denoted by $D(8+6)$ in the following.

Design $D(8+6)$ seems to be rather strange because the groups are treated differently.
An axiomatic criterion of fairness can be \emph{equal treatment of equals} \citep{Palacios-Huerta2012, BramsIsmail2018}, that is, if all clubs are equally strong then each of them should have the same probability of being the final winner even \emph{ex post}, after the drawing of the groups.
However, in the case of format $D(8+6)$:
\begin{itemize}
\item
a team from groups A and B will be the group winner with a probability of $1/8$ and will be eliminated with a probability of $1/4$, thus it qualifies for the quarter-finals with a probability of $1/8 + 5/8 \times 1/2 = 7/16$; while
\item
a team from groups C and D qualifies for the play-off with a probability of $1/3$, for the first knock-out phase with a probability of $1/6$, and for the quarter-finals with a probability of $1/12$. 
\end{itemize}
Hence, a ``lucky'' team has an $84/16 = 5.25$ times higher chance to win the competition than another with the assumption of homogeneous teams.
However, this unrealistic scenario has rather only a theoretical relevance since sports contests are usually unbalanced, which justifies an investigation via Monte-Carlo simulations.

To understand the main characteristics of format $D(8+6)$, two alternative designs will be considered.

First, we have devised a candidate with balanced groups and the same number of teams for comparative purposes. The alternative format $D(4 \times 7)$ is outlined in Figure~\ref{Fig_A2}: the four group winners and the four runners-up qualify directly for the Round of 16, while the third-, fourth-, fifth-, and sixth-placed teams play against each other in the first knock-out phase, where the winners advance to the Round of 16.

Second, the previous format of the EHF Champions League, used between the seasons \href{https://en.wikipedia.org/wiki/2009\%E2\%80\%9310_EHF_Champions_League}{2009/10} and \href{https://en.wikipedia.org/wiki/2014\%E2\%80\%9315_EHF_Champions_League}{2014/15}, is also included in the analysis.\footnote{~We thank an anonymous referee for this suggestion.}
It is shown in Figure~\ref{Fig_A3}: there are four groups of six teams each, the top four clubs qualify for the Round of 16, where a standard knock-out stage starts with drawn brackets---its details will be discussed later. The design is denoted by $D(4 \times 6)$. Note that $D(4 \times 6)$ contains only 24 competitors, which might increase the probability of winning compared to the other two designs, but the latter are organised with more matches and the top teams can skip a knockout round.

\section{Methodology} \label{Sec3}

The three tournaments designs, the innovative $D(8+6)$, as well as the classical $D(4 \times 7)$ and $D(4 \times 6)$ with balanced groups, will be analysed via simulation techniques.

\subsection{The simulation of match outcomes} \label{Sec31}

Most numerical studies of tournament designs apply specific models for simulating match results \citep{ScarfYusofBilbao2009, ScarfYusof2011, GoossensBelienSpieksma2012, LasekGagolewski2018, CoronaForrestTenaWiper2019, DagaevRudyak2019}, but we do not follow this approach due to several reasons.

First, general works comparing different competition formats \citep{Appleton1995, McGarrySchutz1997, Marchand2002} or ranking methods \citep{MendoncaRaghavachari2000} avoid the use of specific prediction models.
Second, while there exists a number of such models for football matches \citep{Maher1982, DixonColes1997, KoningKoolhaasRenesRidder2003, TutzSchauberger2015}, handball seems to be a more difficult sport with respect to forecasting since it is a fast, dynamic, and high-scoring game. Significant differences can be observed between the total number of goals scored per match across the leading men's handball national leagues together with an increasing trend in all countries \citep{MeletakosBayios2010}. Furthermore, the dynamics of handball matches violate independence and identical distribution, sometimes showing a non-stationary behaviour \citep{DumanganeRosatiVolossovitch2009}. Unsurprisingly, \citet{GrollHeinerSchaubergerUhrmeister2019} is probably the first project to forecast handball results through a collaboration of professional statisticians and handball experts.
Third, it is almost impossible to adequately address all issues influencing match outcomes. For instance, the schedule of round-robin tournaments may result in a substantial advantage for some contestants as recent analytical \citep{KrumerMegidishSela2017a, KrumerMegidishSela2020a, Sahm2019} and empirical works \citep{KrumerLechner2017} show. Similarly, even the kick-off time can affect various aspects of games such as the home advantage of the underdog team \citep{Krumer2020}.

Finally, the main message of the current paper is the consideration of competition designs with a non-traditional round-robin group stage. Since this may be relevant in other sports, it makes no sense to fit a particular prediction model on the results of the EHF Champions League matches because it would not contribute much to the general applicability of our suggestion.

Thus, the probability with which a given team would beat another team is fixed \emph{a priori}, and it neither changes during the competition (stationarity) nor is influenced by the previous results (independence). While these conditions clearly do not hold in practice, they can offer a good approximation of long-run averages \citep{McGarrySchutz1997}.
In addition, a variety of models within reason could be taken to determine the winners for comparative purposes \citep{Appleton1995}.

Therefore, each match of a tournament is organised as a Tullock contest such that:
\begin{equation} \label{eq1}
p_{ij} = \frac{(57-i)^r}{(57-i)^r + (57-j)^r}
\end{equation}
gives the probability of team $i$ winning against team $j$, where $r \geq 0$ reflects the discriminatory power of the contest and $1 \leq i,j \leq 28$ is the \emph{identifier} of the teams, which can be called their \emph{pre-tournament rank}.

Consequently, draws are not allowed in any match, although it is not a rare event in handball: in the \href{https://en.wikipedia.org/wiki/2017\%E2\%80\%9318_EHF_Champions_League}{2017/18 EHF Champions League}, there were $10$ in group A, $8$ in group B (from $56$ matches, respectively), as well as $0$ in group C and $3$ in group D (from $30$ matches, respectively).
This assumption is also relatively standard in theoretical papers comparing tournament formats \citep{Appleton1995, McGarrySchutz1997, Marchand2002, Csato2019g}.

The formula $57-i$ has been chosen as the effort of the teams because (1) it should be a non-decreasing function of the pre-tournament rank, and (2) the straightforward solution of $1 \leq 29-i \leq 28$ would lead to a much higher competitive balance among the top teams compared to the competitive balance among the bottom teams, which is implausible.
Throughout the paper, two different values of $r$ ($= 3,5$) will be studied to check the robustness of our results.

Naturally, teams ranked 25--28 are not considered in the design $D(4 \times 6)$.

\input{Figure1_logistic_probability}

Figure~\ref{Fig1} depicts the probability of winning for certain clubs based on formula~\eqref{eq1}.
It can be seen that adjacent teams are closely matched, team $k-1$ defeats team $k$ with a probability of no more than $55$\% even if $r = 5$.
On the other hand, there is a substantial difference between a top club and an underdog: the strongest team has more than $70$\% chance to win against the bottom half of the teams even if $r = 3$.

\subsection{Draws in the group and knock-out stages} \label{Sec32}

Seeding may play a substantial role in knock-out tournaments \citep{Hwang1982, Schwenk2000, Marchand2002, GrohMoldovanuSelaSunde2012, Karpov2016, DagaevSuzdaltsev2018, Karpov2018}.
Although the knock-out stage of competition formats $D(8+6)$ and $D(4 \times 7)$ is predetermined by the previous group stage (see Figures~\ref{Fig_A1}--\ref{Fig_A2}) -- with the exception of the Final Four when there is only one pot --, the clubs should be drawn into groups before the start of the tournament, which may affect its outcome \citep{Guyon2015a, BoczonWilson2018, Guyon2018a, DagaevRudyak2019, LalienaLopez2019, CeaDuranGuajardoSureSiebertZamorano2020}. The knock-out brackets of format $D(4 \times 6)$ are also random.

Compared to the more popular UEFA Champions League, the composition of the pots in the EHF Champions League seems to be less regulated, it depends heavily on the decisions of the EHF Executive Committee \citep{EHF2019b, EHF2019a}. Nevertheless, the administrators obviously intend to place the strongest clubs into groups A and B, including the titleholder, the champions, and the runners-up of the strongest associations, while the champions of low-ranked associations and the runners-up of middle-ranked associations go to groups C and D. Furthermore, groups A and B (C and D) are drawn from eight (six) pots such that the best teams are coming from the first pots.

Therefore, two variants of each tournament design, called \emph{seeded} and \emph{random}, will be examined. They can be considered as two extreme cases: in the real-world, it is impossible to perfectly identify the strength of a team, however, the organiser has some information on it because past performance, the composition of the squad, etc.\ are known before the draw. The actual scenario lies somewhere between these limits, and if a tournament design has more favourable metrics with both assumptions, then it would probably be better in practice, too.

In the seeded version of $D(8+6)$, the two teams having the highest pre-tournament rank are placed in Pot 1, the next two teams are placed in Pot 2, and so on. Groups A and B get a club from each of Pots 1--8 randomly, while groups C and D get a club from each of Pots 9--14 randomly.
Analogously, in the seeded versions of $D(4 \times 7)$ and $D(4 \times 6)$, the four teams with the highest pre-tournament rank are placed in Pot 1, the next four teams are placed in Pot 2, and so on. The four groups get a club from each pot randomly. In the following, these formats are denoted by $D(8+6)/S$, $D(4 \times 7)/S$, and $D(4 \times 6)/S$, respectively.

The random variants work along similar lines but assume some uncertainty in the identification of the teams, while their strength remains unchanged.
In particular, the teams are reranked on the basis of the stochastic values given by the formula $44 \times Rnd + (28-i)$, where $i$ is the teams' pre-tournament rank and $Rnd$ is a random number drawn uniformly from the semi-open range $\left[ 0,1 \right)$. The seeding into pots is based on these estimated coefficients. The appropriate versions are denoted by $D(8+6)/R$, $D(4 \times 7)/R$, and $D(4 \times 6)/R$, respectively.

\input{Figure2_random_seeding}

Figure~\ref{Fig2} shows the probability that a club goes to groups $A$ and $B$ in the design $D(8+6)/R$, in other words, it is classified among the strongest $16$. The best team has more than 88\% chance to achieve this (compared to the 57.14\% of full randomness) and the lowest-ranked team still has around 25\% chance to be drawn into the two top groups.

Finally, there is some randomness in the knock-out stage of the previous EHF Champions League design $D(4 \times 6)$. In its Round of 16, the group winners play against the fourth-placed teams such that teams from the same pot cannot be drawn against each other \citep{EHF2015a}. Contrary to the UEFA Champions League, there is no protection for the teams from the same country. The same rules apply for the runners-up and third-placed teams, which are drawn against each other. In the quarter-finals, the first pot contains the winners of group winners versus fourth-placed teams, and the second pot contains the winners of runners-up versus third-placed teams \citep{EHF2015b}. There are no restrictions, even the winner and the runner-up of the same group can meet as happened in the \href{https://en.wikipedia.org/wiki/2012\%E2\%80\%9313_EHF_Champions_League_knockout_stage#Quarterfinals}{2012/13} season of the EHF Champions League for Groups A, B, and C.

\begin{table}[htbp]
\centering
\caption{Feasible matchings in the Round of 16 draw of design $D(4 \times 6)$}
\label{Table1}
\rowcolors{1}{gray!20}{}
    \begin{tabularx}{\textwidth}{cCCCC CCCCC} \toprule \hiderowcolors
    Group of the first team & \multicolumn{9}{c}{Group of the second team} \\ \midrule \showrowcolors
    \textbf{A} & B     & B     & B     & C     & C     & C     & D     & D     & D \\
    \textbf{B} & A     & C     & D     & A     & D     & D     & A     & C     & C \\
    \textbf{C} & D     & D     & A     & D     & A     & B     & B     & A     & B \\
    \textbf{D} & C     & A     & C     & B     & B     & A     & C     & B     & A \\ \toprule
    \end{tabularx}
\end{table}

In the Round of 16 draw, there are $4! = 24$ possible scenarios to pair the groups. Among them, $15$ violates the restriction, hence we chose from the remaining 9 matchings with a uniform probability of $1/9$ for the four clashes between group winners and fourth-placed teams, as well as for the four matches between runners-up and third-placed teams. They are listed in Table~\ref{Table1}. In the absence of any restrictions, the draw of the quarter-finals is straightforward after the pots are determined. 

\subsection{Tournament metrics and details of the simulation procedure} \label{Sec33}

Choosing a particular design and a prediction model for match outcomes, the competition can be simulated repeatedly in order to obtain any metrics of interest.

We will analyse the following success measures, which are widely used in evaluating tournament designs via simulations \citep{ScarfYusofBilbao2009, DagaevRudyak2019}:
\begin{itemize}
\item
the average pre-tournament ranks of the clubs in the Final Four, that is, the winner, the second-, third-, and fourth-placed teams;
\item
the expected quality of all matches, measured by the sum of the playing teams' pre-tournament ranks;
\item
the expected competitive balance of all matches, measured by the difference between the playing teams' pre-tournament ranks.
\end{itemize}
In the case of the first metric, our focus is on the first four places because the Final Four of the Champions League is promoted by the EHF separately. For example, this event has its own website (\url{http://www.ehffinal4.com/}), and tickets can be bought for the whole weekend, which offers an ``\emph{indisputable highlight of the European club handball season}'', instead of the individual matches \citep{EHF2019c}.

Hence, the first four clubs, as well as the number of matches played by any two clubs, and the winning percentage of each club have been recorded.
Note also that a lower value is preferred for all tournament metrics.

According to Section~\ref{Sec31}, draws between the teams are not allowed in any match.
This is not to be confused with ties in the ranking of round-robin groups, resolved in our simulations with an equal-odds coin toss.
Furthermore, both formats contain a knock-out stage with home-away matches before the Final Four. If one team wins the first and the other wins the second match, then the qualifying team is chosen randomly with the probability given by formula~\eqref{eq1}, which is equivalent to the assumption that the clash is decided by three matches.
It is used because resolution by a pure coin toss would be an inappropriate solution in the case of a fixed winning probability $p$: the chance to qualify with two matches is $p^2 + 2 \times p(1-p) \times 0.5 = p$ as the probability of a tied contest is $2 \times p(1-p)$, therefore, a knock-out played over two legs would be the same as a knock-out played over one leg.

\input{Figure3_number_of_runs}

To get a reasonable estimate of all tournament metrics despite the stochastic nature of the simulations, we have determined the required number of independent runs on the basis of the random variant of design $D(8+6)$. Figure~\ref{Fig3} shows that both success measures analysed for this purpose, the proportion of tournament wins for the strongest club, and the average number of matches between the two strongest clubs in one iteration, remain practically unchanged after one million ($10^6$) iterations. Since, in the view of model limitations, there is no need to further reduce the statistical error, it has been decided that all subsequent simulations will be implemented with one million runs ($N=1{,}000{,}000$).

The validity of the simulation procedure has been investigated in several ways.
First, the assumption of equally strong teams ($p_{ij} = 0.5$ for all combinations of $i$ and $j$) has led to, as expected, an outcome where all teams are placed first to fourth equally often in format $D(4 \times 7)$, but not in $D(8+6)$, see Section~\ref{Sec2}. However, the chances of the first $16$ and the last $12$ clubs are the same according to design $D(8+6)/S$.
Second, we have analysed a fully deterministic matrix ($p_{ij} = 1$ if $i < j$), which implies that the top team wins, the fourth team is the fourth, while the second strongest occupies the second position with a probability of $2/3$ (when it does not play against the best team in the semi-final) and the third position with a probability of $1/3$ (when it plays against the best team in the semi-final) in the seeded variant of all tournament formats. 
Finally, changing certain values of the fully deterministic matrix has been checked to modify the outcome concerning the first four places accordingly.

\subsection{The main limitations of our approach} \label{Sec34}

No theoretical model can be a perfect substitute for real-world experiments. Any interpretation of our results should be careful due to the following reasons:
\begin{itemize}
\item
The strength of all teams is exogenously given and fixed during the whole tournament.
\item
Draws are not allowed, which is not in line with the rules of handball.
\item
Home advantage is disregarded both in the group and elimination stages. This can be a problem if playing the second leg at home means an advantage. In addition, German teams might benefit from organising the Final Four in Cologne. 
\item
It is assumed that teams exert full effort in all games. This is not necessarily true, especially if a team has already qualified from its group. In particular, according to the theoretical result of \citet{Vong2017}, if more than one team qualify from a group, a team may potentially benefit from exerting zero effort in some matches even when this is costless.
\item
Goal difference is taken into account neither in the group stage nor in the knock-out phase. However, the choice of the winner when both teams have won one match each in the elimination stage favours the better team, which somewhat mitigates this problem.
\end{itemize}
To conclude, the use of a particular probabilistic model always implies certain limitations. Nonetheless, some efforts will be made to minimise this weakness by carrying out robustness checks with respect to competitive balance and seeding, and a wide range of models may be suitable for comparative purposes \citep{Appleton1995}.

\section{Results} \label{Sec4}

In the following, a detailed analysis of the three competition formats is provided.

\begin{table}[ht]
\centering
\caption{The number of matches in a season}
\label{Table2}
\rowcolors{1}{gray!20}{}
    \begin{tabularx}{0.8\textwidth}{lCCC} \toprule \hiderowcolors
    Tournament format & $D(8+6)$ & $D(4 \times 7)$ & $D(4 \times 6)$ \\ \midrule \showrowcolors
    Groups A and B & 112   & 84    & 60 \\
    Groups C and D & 60    & 84    & 60 \\
    Play-off & 4     & ---   & --- \\
    First knock-out phase & 12    & 16    & --- \\
    Round of 16 & ---   & 16    & 16 \\
    Quarter-finals & 8     & 8     & 8 \\
    Final Four & 4     & 4     & 4 \\ \hline
    \textbf{Total} & \textbf{200}   & \textbf{212}   & \textbf{148} \\ \toprule
    \end{tabularx}
\end{table}

The tournament designs differ in the number of matches played according to Table~\ref{Table2}. The innovative format $D(8+6)$ requires fewer matches than its traditional pair $D(4 \times 7)$. While $D(4 \times 6)$ contains only $24$ teams, the reduction in the number of matches is more substantial.

\input{Figure4_match_number}

Figure~\ref{Fig4} presents the distribution of the matches, the number of teams playing a given number of matches supposed that a stronger team always beats a weaker one (otherwise, the distribution may change). For example, in the design $D(4 \times 7)$, the four teams eliminated after the group stage play $12$ matches each.
The variance of the number of matches is the greatest in format $D(8+6)$.

\input{Figure5_average_rank}

Our first contest metric is the average pre-tournament ranks of the participants in the Final Four. It is revealed by Figure~\ref{Fig5} that the seeded design $D(8+6)/S$ is superior in the ability to select the strongest teams, followed by its random variant $D(8+6)/R$, while the two versions of the competition formats $D(4 \times 7)$ and $D(4 \times 6)$ are close to each other, the seeded being somewhat more efficacious. This finding is robust across the four places and does not depend on the variation in the strengths of the clubs as the diagrams are similar for all values of the competitiveness parameter $r$.

\input{Figure6_quality_competitive_balance}

The picture is somewhat different for the other two tournament success measures, the expected quality and competitive balance of the matches (Figure~\ref{Fig6}). Remember that a lower value of quality means that the matches are played by stronger teams on average. Therefore, since the format $D(4 \times 6)$ does not contain the four weakest teams, it proves to be the best for this metric. On the other hand, design $D(8+6)$ outperforms $D(4 \times 7)$, even if the teams are not allocated perfectly into the top and bottom groups.

Analogously, a smaller competitive balance corresponds to more uncertainty in the outcome of a match. Now the random variant of formats $D(4 \times 7)$ and $D(4 \times 6)$ has a more favourable value compared to the seeded versions because the optimal seeding guarantees that the strong teams play against underdogs in the groups. While competitive balance substantially deteriorates for design $D(8+6)$ when the clubs cannot be allocated perfectly into the two types of groups, $D(8+6)/R$ is still consistently better than the alternative design $D(4 \times 7)$, and not worse than $D(4 \times 6)$, which is organised with four teams less.

It is worth recalling here that unbalanced groups are recommended to raise the expected quality and competitive balance of all matches played, thus the more visible and robust advantage of the format $D(8+6)$ over $D(4 \times 7)$ in Figure~\ref{Fig6} compared to Figure~\ref{Fig5} reinforces our main message. In fact, a substantial increase in match quality and competitive balance can be probably tolerated even if the selection ability of the championship, that is, the average pre-tournament ranks of the contestants in the Final Four would somewhat worsen.

The structure of format $D(8+6)$ opens the possibility that a team might benefit from being in the bottom groups C or D instead of the top groups A and B.
This is not only an academic problem, for instance, the \href{https://en.wikipedia.org/wiki/2017\%E2\%80\%9318_EHF_Champions_League}{2017/18 EHF Champions League} was won by \href{https://en.wikipedia.org/wiki/Montpellier_Handball}{Montpellier Handball}, a club which started from group C after being the third in the \href{https://en.wikipedia.org/wiki/2016\%E2\%80\%9317_LNH_Division_1\#League_table}{previous season} of the \href{https://en.wikipedia.org/wiki/LNH_Division_1}{LNH (\emph{Ligue Nationale de Handball}) Division 1}, the French premier handball league.

If a strong team gains from masking itself as a weaker one, this can be called unfair because it gives incentives to tank. For example, the qualification for the UEFA European Championship 2020 is demonstrated to exhibit such a shortcoming \citep{Csato2020b}.

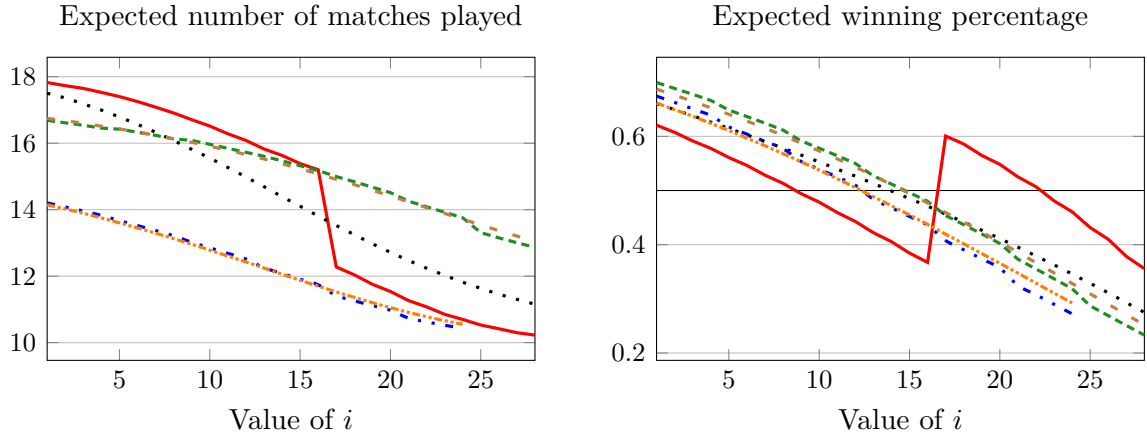
Hence an alternative scenario called \emph{erroneous team identification} is considered when:
\begin{itemize}
\item
the $k$th pre-tournament ranked team is correctly identified as the $k$th strongest one by the seeding procedure for all $1 \leq k \leq 8$ and $18 \leq k \leq 28$;
\item
the $9$th pre-tournament ranked team is identified as the $17$th strongest one by the seeding procedure;
\item
the $\ell$th pre-tournament ranked team is identified as the $(\ell -1)$th strongest one by the seeding procedure for all $10 \leq \ell \leq 17$.
\end{itemize}
To be short, contrary to the original case of \emph{correct team identification}, the $9$th best club now seems to be only the $17$th before the tournament, therefore it might obtain a less difficult path into the Final Four.
Note that, while the current quality of the teams is obviously unknown in practice, the random variants of the tournament formats also address this kind of uncertainty.

\input{Figure7_matches_points}

Figure~\ref{Fig7} shows the average number of matches played and the average winning performance for each team under both scenarios. There is a break in both measures between the $16$th and $17$th strongest teams if the seeded $D(8+6)/S$ design is used, but it is smoothed out by the random variant $D(8+6)/R$.
Format $D(8+6)/R$ involves more matches between the leading teams and reduces the variance in winning probabilities compared to the traditional design of $D(4 \times 7)$. Differences in the winning percentage are greater for the random versions of designs $D(4 \times 7)$ and $D(4 \times 6)$ than for the seeded versions because weak teams are forced to play against strong teams by the latter regime.

According to Figure~\ref{Fig7b}, the $9$th team, which is erroneously identified as an underdog, plays fewer matches with a higher winning percentage as it becomes the best team of the bottom groups in the competition design $D(8+6)/S$. However, the random variant $D(8+6)/R$ substantially weakens the effect of this error.

\input{Figure8_expected_prize}

Fairness is highlighted in Figure~\ref{Fig8} by plotting the ratio of expected prizes between a club and the best club among the clubs that are weaker than it. The expected prize is defined here by giving five points to the tournament winner, three points to the second-, two points to the third-, and one point to the fourth-placed team. The ratio is not calculated for the last four teams because the expected prize of low-quality clubs becomes volatile.

In the case of correct team identification, these ratios are consistently over one for low values of the competitiveness parameter $r$, that is, in a more competitive tournament (Figure~\ref{Fig8a}). On the other hand, the $17$th club has a higher chance to achieve a good position than the $16$th club in the format $D(8+6)/S$ if the teams' a priori strength differ significantly ($r=5$). However, this seems to be only a marginal violation of fairness as the expected prize of the $17$th team is still lower than the expected prize of the $15$th, and introducing randomness into the identification of the teams immediately solves the problem.

The potential unfairness caused by erroneous team identification is even more mitigated (Figure~\ref{Fig8b}). In particular, the $9$th club loses by being listed among the underdogs as it obtains a lower expected prize than the $10$th club in the seeded design $D(8+6)/S$ if $r=3$. Furthermore, random seeding ($D(8+6)/R$) or more diverse pre-tournament strength of the teams ($r=5$) eliminates the possibility that a weaker team goes into the Final Four with a higher probability than a stronger one.
The existence of top and bottom groups in design $D(8+6)$ did not contribute to the unexpected victory of Montpellier.

According to Table~\ref{Table_A1} in the Appendix, erroneous team identification does not influence our tournaments metrics, the efficacy of the average ranks of the first four-placed teams, the expected quality of all matches, and outcome uncertainty, in a meaningful way. It also reinforces that the choice of the tournament format is more important than the correct identification of the teams before the seeding. 
Furthermore, format $D(8+6)$ remains undoubtedly superior to $D(4 \times 7)$ despite that fewer matches are played. On the other hand, this implication would not be true concerning $D(4 \times 6)$ as selecting the best teams for the Final Four is simpler on the basis of more matches. 


\section{Conclusions} \label{Sec6}

The design of a hybrid tournament consisting of round-robin groups followed by a knock-out phase raises several interesting theoretical questions on various fields such as fairness \citep{Guyon2018a, CeaDuranGuajardoSureSiebertZamorano2020}, or strategy-proofness \citep{Pauly2014, Vong2017, DagaevSonin2018, Csato2019b, Csato2020c}.
The current paper has attempted to explore the potential effects of creating groups with different quality through the example of the EHF Champions League, the most prestigious men's handball club competition in Europe. Its non-traditional design for $28$ teams, applied since the 2015/16 season, has been compared to the classical format of four balanced, equally strong groups with seven teams each, and to the previous design of the tournament.

Compared to the formats containing balanced groups, the innovative competition design is able to considerably increase the proportion of high quality and even matches, with positive effects on demand. In addition, it turns out to be more \emph{efficacious} as the average pre-tournament ranks of the teams finishing at any positions in the Final Four are smaller.
Our numerical results have revealed the new competition design is in line with the intentions of the EHF \citep{EHF2014c}, too.
Nonetheless, both the men's and the women's EHF Champions League competitions will be played with two balanced groups of eight teams from the 2020/21 season \citep{EHF2018}.

It should be recognised that there are always some unavoidable trade-offs between the goals of a tournament designer, and they cannot be achieved simultaneously as recent theoretical results show \citep{KrumerMegidishSela2017b}. For example, while format $D(8+6)$ is better at selecting stronger teams to be the winner, decreasing the chances that a weaker team plays against a stronger one can be detrimental for gradually improving the squad of the former. Furthermore, bottom groups without leading clubs can generate limited media attention.

The idea of unbalanced groups may be used in several other settings.
UEFA seems to follow this principle by the structure of the new biennial international football competition \href{https://en.wikipedia.org/wiki/UEFA_Nations_League}{UEFA Nations League}, which started in 2018: the 55 UEFA national teams are divided into four divisions called leagues such that in each of them, four groups are formed with teams of similar quality, achieved by promotion and relegation between the leagues over time.
An alternative design can also be considered for the UEFA Champions League with the first three teams from four top groups and the group winners from four bottom groups qualifying for the Round of 16. This solution could create more matchups between the richest clubs without imposing severe barriers to entry for teams outside the current elite.

To summarise, we have successfully challenged the traditional seeding system used in hybrid tournaments. It is worth considering an alternative format composed of unbalanced groups with more even matchings within groups, which are treated unequally to guarantee an easier path for teams from the top groups to qualify for the next stage.

\section*{Acknowledgements}
\addcontentsline{toc}{section}{Acknowledgements}
\noindent
\emph{My father} (also called \emph{L\'aszl\'o Csat\'o}) has made a substantial contribution to the paper by helping to code the simulations in Python. \\
We are grateful to \emph{Tam\'as Halm} for reading the manuscript. \\
Five anonymous reviewers provided valuable comments and suggestions on earlier drafts. \\
We are indebted to the \href{https://en.wikipedia.org/wiki/Wikipedia_community}{Wikipedia community} for collecting and structuring valuable information used in our research. \\
The research was supported by the MTA Premium Postdoctoral Research Program grant PPD2019-9/2019.

\bibliographystyle{apalike}
\bibliography{All_references}

\clearpage

\section*{Appendix}
\addcontentsline{toc}{section}{Appendix}

\renewcommand\thetable{A.\arabic{table}}
\setcounter{table}{0}

\makeatletter
\renewcommand\p@subtable{A.\arabic{table}}
\makeatother

\renewcommand\thefigure{A.\arabic{figure}}
\setcounter{figure}{0}

\makeatletter
\renewcommand\p@subfigure{A.\arabic{figure}}
\makeatother

\begin{table}[ht!]
\centering
\caption[Estimates of tournament metrics for all designs]{Estimates of tournament metrics for all designs \\
\small{$D86$, $D77$, and $D66$ stand for competition designs $D(8+6)$, $D(4 \times 7)$, and $D(4 \times 6)$, respectively}}
\label{Table_A1}

\small{
\begin{subtable}{\textwidth}
\caption{$r=3$ (more competitive), correct team identification}
\rowcolors{1}{gray!20}{}
    \begin{tabularx}{1\textwidth}{m{4cm} RRR RRR} \toprule \hiderowcolors
    	& \multicolumn{1}{c}{D86/S} & \multicolumn{1}{c}{D86/R} & \multicolumn{1}{c}{D77/S} & \multicolumn{1}{c}{D77/R} & \multicolumn{1}{c}{D66/S} & \multicolumn{1}{c}{D66/R} \\ \midrule \showrowcolors
    Average rank of \#1 & 4.959 & 5.025 & 5.334 & 5.338 & 5.352 & 5.397 \\
    Average rank of \#2 & 5.951 & 6.073 & 6.520 & 6.530 & 6.531 & 6.622 \\
    Average rank of \#3 & 5.968 & 6.092 & 6.543 & 6.561 & 6.563 & 6.653 \\
    Average rank of \#4 & 7.279 & 7.508 & 8.119 & 8.142 & 8.147 & 8.329 \\ \hline
    Expected quality of \newline all matches & 48.80 & 50.34 & 55.52 & 55.57 & 47.27 & 47.44 \\
    Expected competitive \newline balance of all matches & 10.67 & 16.43 & 20.24 & 19.02 & 17.82 & 16.48 \\ \toprule
    \end{tabularx}
\end{subtable}

\vspace{0.25cm}
\begin{subtable}{\textwidth}
\caption{$r=3$ (more competitive), erroneous team identification}
\rowcolors{1}{gray!20}{}
    \begin{tabularx}{1\textwidth}{m{4cm} RRR RRR} \toprule \hiderowcolors
    	& \multicolumn{1}{c}{D86/S} & \multicolumn{1}{c}{D86/R} & \multicolumn{1}{c}{D77/S} & \multicolumn{1}{c}{D77/R} & \multicolumn{1}{c}{D66/S} & \multicolumn{1}{c}{D66/R} \\ \midrule \showrowcolors
    Average rank of \#1 & 4.951 & 5.021 & 5.338 & 5.340 & 5.349 & 5.394 \\
    Average rank of \#2 & 5.946 & 6.088 & 6.528 & 6.531 & 6.540 & 6.631 \\
    Average rank of \#3 & 5.963 & 6.106 & 6.551 & 6.562 & 6.577 & 6.666 \\
    Average rank of \#4 & 7.277 & 7.536 & 8.136 & 8.140 & 8.184 & 8.328 \\ \hline
    Expected quality of \newline all matches & 48.75 & 50.40 & 55.57 & 55.61 & 47.37 & 47.53 \\
    Expected competitive \newline balance of all matches & 10.67 & 16.46 & 20.27 & 19.05 & 17.88 & 16.53 \\
\toprule
    \end{tabularx}
\end{subtable}

\vspace{0.25cm}
\begin{subtable}{\textwidth}
\caption{$r=5$ (less competitive), correct team identification}
\rowcolors{1}{gray!20}{}
    \begin{tabularx}{1\textwidth}{m{4cm} RRR RRR} \toprule \hiderowcolors
    	& \multicolumn{1}{c}{D86/S} & \multicolumn{1}{c}{D86/R} & \multicolumn{1}{c}{D77/S} & \multicolumn{1}{c}{D77/R} & \multicolumn{1}{c}{D66/S} & \multicolumn{1}{c}{D66/R} \\ \midrule \showrowcolors
    Average rank of \#1 & 3.786 & 3.891 & 4.052 & 4.059 & 4.028 & 4.075 \\
    Average rank of \#2 & 4.707 & 4.917 & 5.169 & 5.172 & 5.117 & 5.202 \\
    Average rank of \#3 & 4.747 & 4.946 & 5.192 & 5.205 & 5.159 & 5.244 \\
    Average rank of \#4 & 6.072 & 6.391 & 6.776 & 6.815 & 6.714 & 6.913 \\ \hline
    Expected quality of \newline all matches & 48.41 & 49.89 & 55.04 & 55.09 & 46.52 & 46.72 \\
    Expected competitive \newline balance of all matches & 10.54 & 16.24 & 20.03 & 18.80 & 17.62 & 16.29 \\ \toprule
    \end{tabularx}
\end{subtable}

\vspace{0.25cm}
\begin{subtable}{\textwidth}
\caption{$r=5$ (less competitive), erroneous team identification}
\rowcolors{1}{gray!20}{}
    \begin{tabularx}{1\textwidth}{m{4cm} RRR RRR} \toprule \hiderowcolors
    	& \multicolumn{1}{c}{D86/S} & \multicolumn{1}{c}{D86/R} & \multicolumn{1}{c}{D77/S} & \multicolumn{1}{c}{D77/R} & \multicolumn{1}{c}{D66/S} & \multicolumn{1}{c}{D66/R} \\ \midrule \showrowcolors
    Average rank of \#1 & 3.801 & 3.885 & 4.055 & 4.052 & 4.018 & 4.072 \\
    Average rank of \#2 & 4.736 & 4.912 & 5.164 & 5.165 & 5.108 & 5.201 \\
    Average rank of \#3 & 4.765 & 4.942 & 5.203 & 5.210 & 5.152 & 5.253 \\
    Average rank of \#4 & 6.098 & 6.410 & 6.781 & 6.815 & 6.727 & 6.910 \\ \hline
    Expected quality of \newline all matches & 48.39 & 49.97 & 55.10 & 55.14 & 46.67 & 46.84 \\
    Expected competitive \newline balance of all matches & 10.56 & 16.29 & 20.06 & 18.84 & 17.72 & 16.36 \\ \toprule
    \end{tabularx}
\end{subtable}
}
\end{table}

\input{Figures_Appendix}

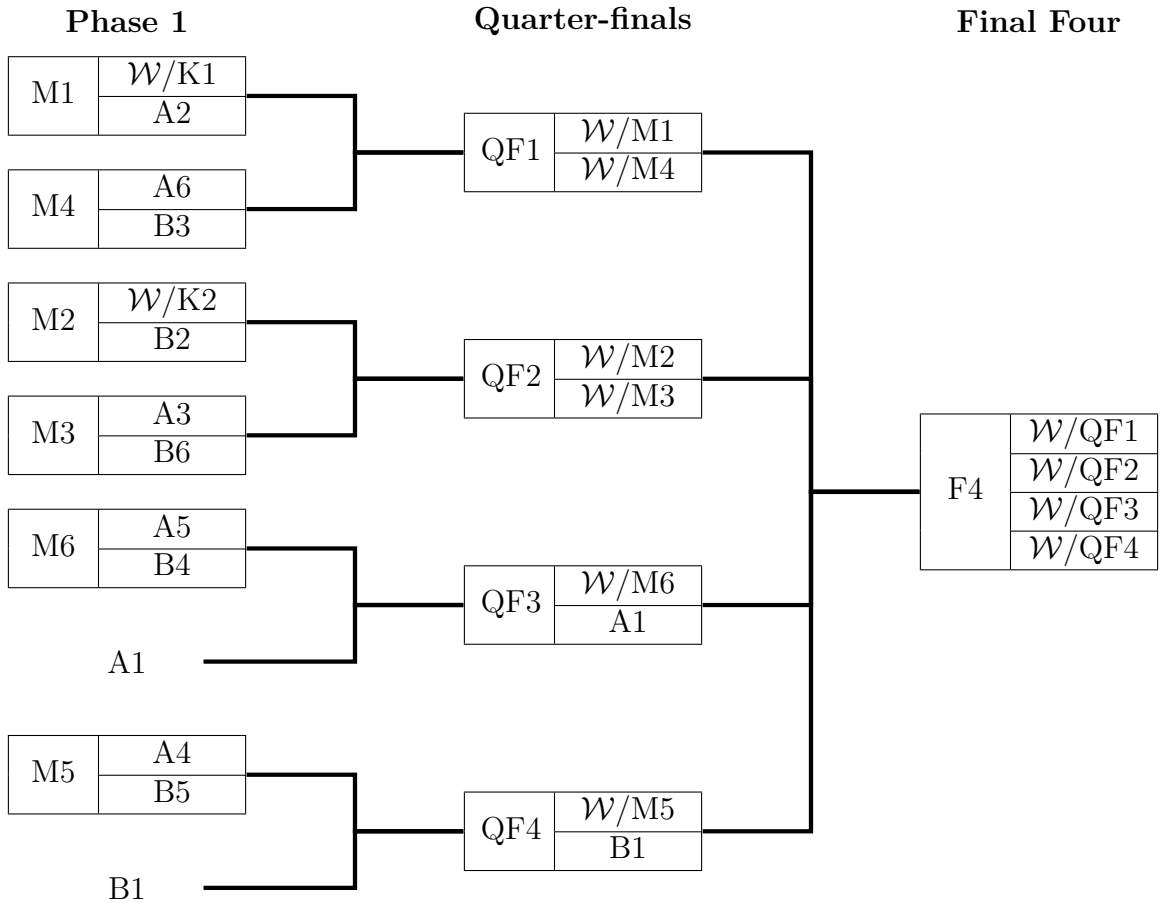
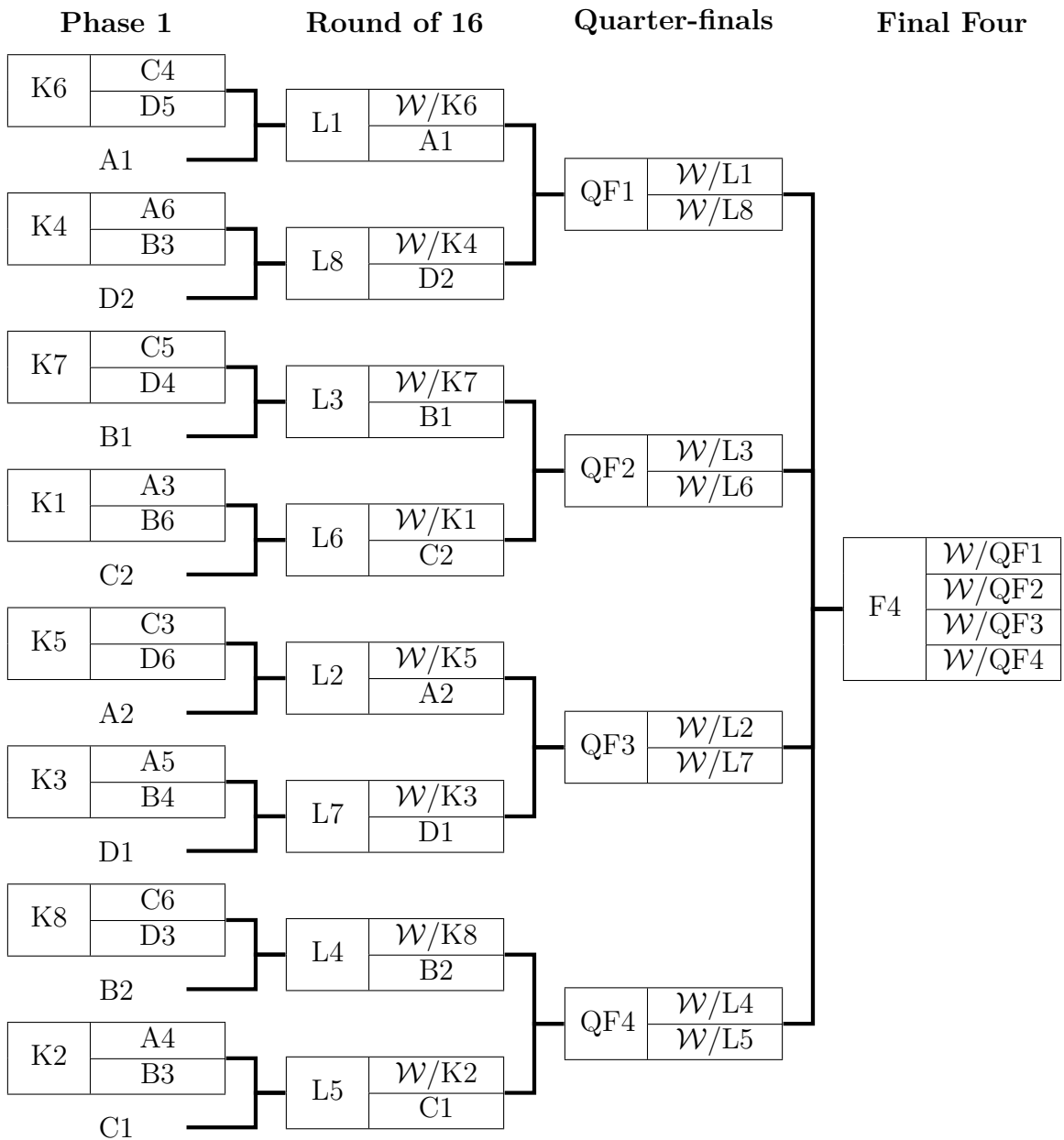
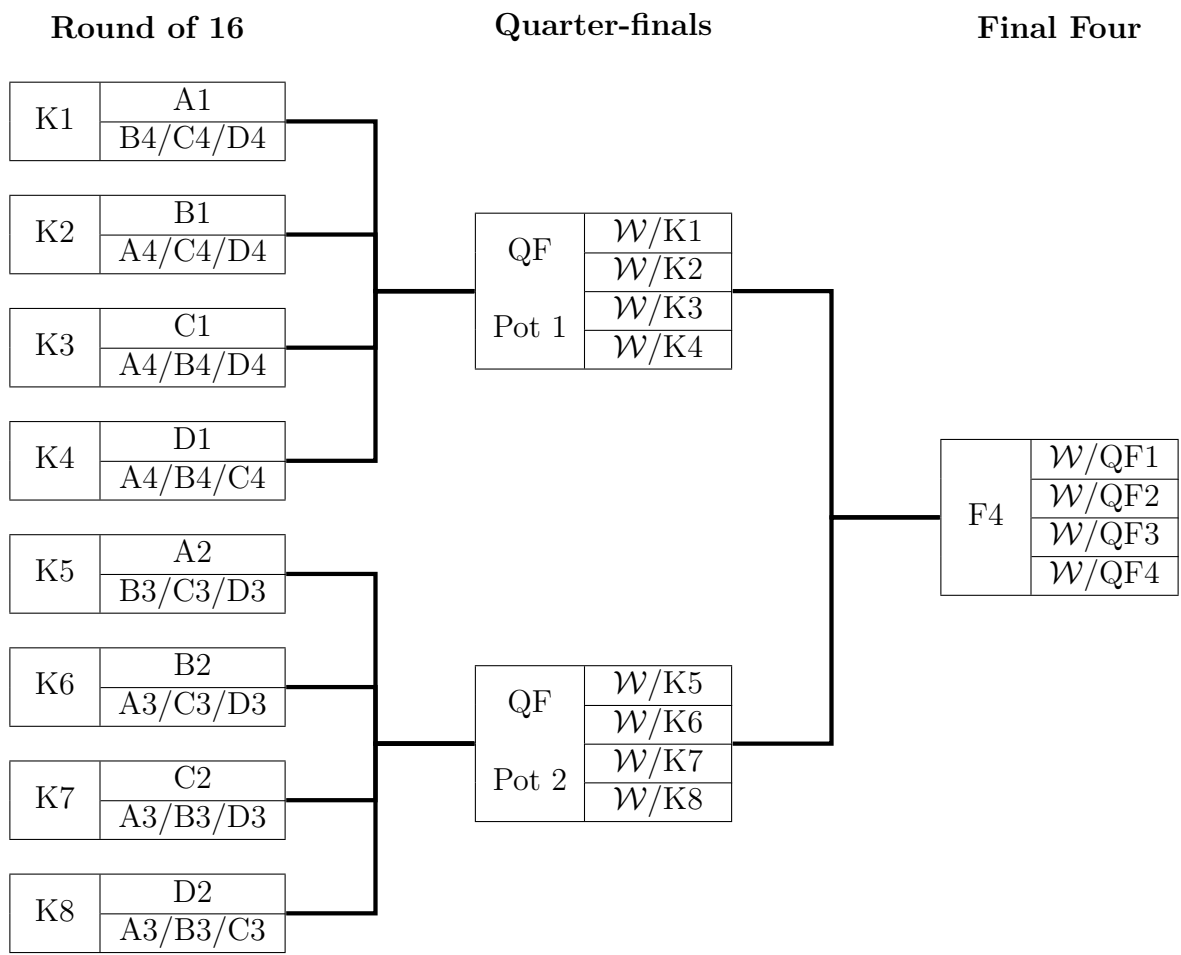
\end{document}

%% file: Figure1_logistic_probability.tex
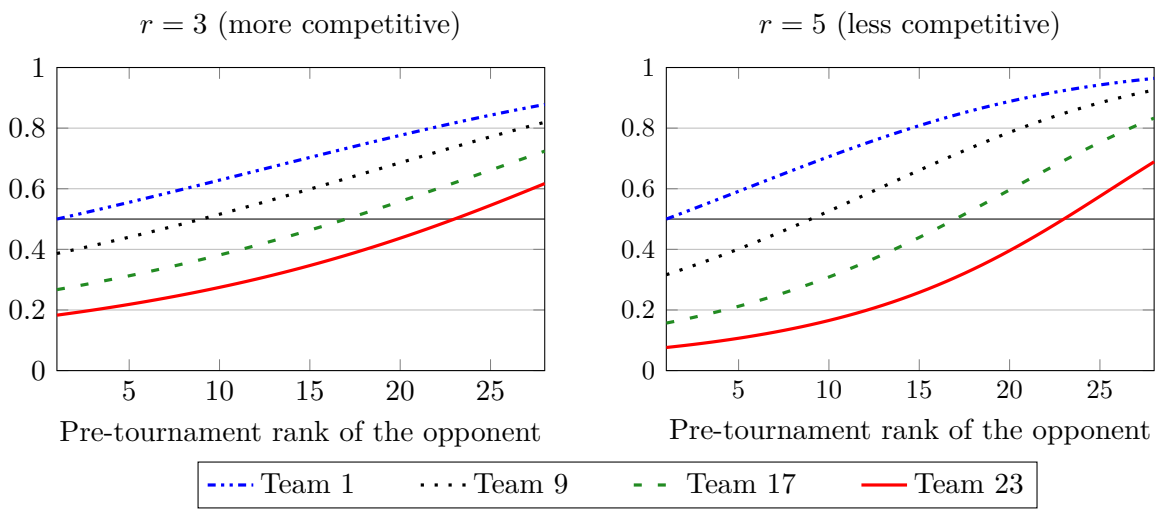
\begin{figure}[ht]
\centering
\caption{The probability that team $i$ beats its opponent}
\label{Fig1}

\begin{tikzpicture}
\begin{axis}[
name = axis1,
width = 0.5\textwidth, 
height = 0.35\textwidth,
title = {$r=3$ (more competitive)},
title style = {align=center, font=\small},
xmin = 1,
xmax = 28,
ymin = 0,
ymax = 1,
ymajorgrids,
xlabel = Pre-tournament rank of the opponent,
xlabel style = {font=\small},
tick label style = {/pgf/number format/fixed, font=\small},
]
\draw (axis cs:\pgfkeysvalueof{/pgfplots/xmin},0.5)  -- (axis cs:\pgfkeysvalueof{/pgfplots/xmax},0.5);
\addplot[blue,smooth,very thick,dashdotdotted] coordinates {
(1,0.5)
(2,0.513510589459957)
(3,0.527248709018854)
(4,0.541201196944156)
(5,0.555353167375025)
(6,0.569687965302806)
(7,0.584187135747931)
(8,0.598830409356725)
(9,0.613595706618962)
(10,0.628459162822655)
(11,0.64339517570855)
(12,0.658376477556881)
(13,0.673374233128834)
(14,0.688358164493205)
(15,0.703296703296703)
(16,0.718157170489537)
(17,0.732905982905983)
(18,0.747508885436397)
(19,0.761931206830724)
(20,0.776138136465888)
(21,0.790095018715808)
(22,0.803767660910518)
(23,0.817122650288479)
(24,0.830127674861619)
(25,0.842751842751843)
(26,0.854965994342939)
(27,0.866743001539859)
(28,0.878058048548786)
};
\addplot[black,smooth,very thick,loosely dotted] coordinates {
(1,0.386404293381038)
(2,0.399296667111966)
(3,0.412570507655117)
(4,0.426224327376295)
(5,0.440254777070064)
(6,0.454656454656455)
(7,0.469421712112466)
(8,0.484540463807992)
(9,0.5)
(10,0.515784809831402)
(11,0.53187641876034)
(12,0.548253245884085)
(13,0.564890487087283)
(14,0.581760030300002)
(15,0.598830409356725)
(16,0.616066802961345)
(17,0.633431085043988)
(18,0.650881932305737)
(19,0.668374993956389)
(20,0.685863127538838)
(21,0.703296703296703)
(22,0.720623977793271)
(23,0.737791535464589)
(24,0.75474479454579)
(25,0.771428571428571)
(26,0.787787695091286)
(27,0.803767660910518)
(28,0.819315311043777)
};
\addplot[ForestGreen,smooth,very thick,loosely dashed] coordinates {
(1,0.267094017094017)
(2,0.277807921866522)
(3,0.288986020301268)
(4,0.300643094369049)
(5,0.312793243665937)
(6,0.325449654464)
(7,0.338624338624339)
(8,0.352327841056103)
(9,0.366568914956012)
(10,0.381354164804586)
(11,0.396687658055239)
(12,0.412570507655117)
(13,0.429000429000429)
(14,0.445971276662462)
(15,0.463472568217369)
(16,0.481489004747181)
(17,0.5)
(18,0.518979232721641)
(19,0.538394239181641)
(20,0.558206065257778)
(21,0.578368999421631)
(22,0.598830409356725)
(23,0.619530705490591)
(24,0.640403454176131)
(25,0.661375661375661)
(26,0.682368244287831)
(27,0.703296703296703)
(28,0.724071999909491)
};
\addplot[red,smooth,very thick] coordinates {
(1,0.182877349711521)
(2,0.191093889021242)
(3,0.199747926492113)
(4,0.208862743847679)
(5,0.218462359375695)
(6,0.228571428571429)
(7,0.239215113448242)
(8,0.250418915216657)
(9,0.262208464535411)
(10,0.274609263102)
(11,0.287646370023419)
(12,0.30134402625183)
(13,0.315725210462053)
(14,0.330811120182475)
(15,0.346620572879921)
(16,0.363169323169323)
(17,0.380469294509409)
(18,0.398527726798009)
(19,0.417346245327897)
(20,0.436919861711707)
(21,0.457235923685435)
(22,0.478273038124095)
(23,0.5)
(24,0.522374769075371)
(25,0.545343545343545)
(26,0.568840002894565)
(27,0.592784749034749)
(28,0.617085079993092)
};
\end{axis}

\begin{axis}[
at = {(axis1.south east)},
xshift = 0.1\textwidth,
width = 0.5\textwidth, 
height = 0.35\textwidth,
title = {$r=5$ (less competitive)},
title style = {align=center, font=\small},
xmin = 1,
xmax = 28,
ymin = 0,
ymax = 1,
ymajorgrids,
xlabel = Pre-tournament rank of the opponent,
xlabel style = {font=\small},
tick label style = {/pgf/number format/fixed},
legend entries={Team $1 \qquad$,Team $9 \qquad$,Team $17 \qquad$,Team $23$},
legend style = {at={(-0.1,-0.3)},anchor=north,legend columns = 4,font=\small}
]
\draw (axis cs:\pgfkeysvalueof{/pgfplots/xmin},0.5)  -- (axis cs:\pgfkeysvalueof{/pgfplots/xmax},0.5);
\addplot[blue,smooth,very thick,dashdotdotted] coordinates {
(1,0.5)
(2,0.522507909843214)
(3,0.545334707467139)
(4,0.568393308373283)
(5,0.591589421383606)
(6,0.614822724689605)
(7,0.637988303155328)
(8,0.660978315683308)
(9,0.683683846560631)
(10,0.705996881335206)
(11,0.727812337199875)
(12,0.749030071204713)
(13,0.769556787694509)
(14,0.789307769572618)
(15,0.808208366219416)
(16,0.826195183535391)
(17,0.843216937587798)
(18,0.859234951355278)
(19,0.874223292579618)
(20,0.888168568273901)
(21,0.901069406724714)
(22,0.912935669907781)
(23,0.923787447548272)
(24,0.933653888440094)
(25,0.942571925298637)
(26,0.950584946851274)
(27,0.957741465751625)
(28,0.964093824006661)
};
\addplot[black,smooth,very thick,loosely dotted] coordinates {
(1,0.316316153439369)
(2,0.336113818861346)
(3,0.356883801474672)
(4,0.378609468158252)
(5,0.401261036081435)
(6,0.424794191294639)
(7,0.449148926100936)
(8,0.474248695906657)
(9,0.5)
(10,0.526292486728904)
(11,0.552999670045775)
(12,0.579980320263104)
(13,0.607080557134597)
(14,0.63413662949301)
(15,0.660978315683308)
(16,0.687432827453332)
(17,0.713329052380516)
(18,0.738501932220508)
(19,0.762796752054825)
(20,0.786073111527708)
(21,0.808208366219416)
(22,0.829100363030648)
(23,0.848669344494276)
(24,0.866858957348009)
(25,0.883636363636364)
(26,0.898991511291946)
(27,0.912935669907781)
(28,0.9254993724155)
};
\addplot[ForestGreen,smooth,very thick,loosely dashed] coordinates {
(1,0.156783062412201)
(2,0.169064952352453)
(3,0.182347538795436)
(4,0.196697822737394)
(5,0.212182230586917)
(6,0.228865044543496)
(7,0.246806459387804)
(8,0.266060237092565)
(9,0.286670947619484)
(10,0.308670809927216)
(11,0.332076182842392)
(12,0.356883801474672)
(13,0.383066910297222)
(14,0.41057150579301)
(15,0.439312963947095)
(16,0.469173382157421)
(17,0.5)
(18,0.531605065882714)
(19,0.563767474809724)
(20,0.596236407898765)
(21,0.628737055875862)
(22,0.660978315683308)
(23,0.692662132182879)
(24,0.723493946139677)
(25,0.753193540612196)
(26,0.781505483462989)
(27,0.808208366219416)
(28,0.833122144192143)
};
\addplot[red,smooth,very thick] coordinates {
(1,0.0762125524517283)
(2,0.0828025963028901)
(3,0.0900423774494945)
(4,0.0979991245924611)
(5,0.106746269312025)
(6,0.116363636363636)
(7,0.126937489148881)
(8,0.138560369457691)
(9,0.151330655505724)
(10,0.165351745881787)
(11,0.180730760418523)
(12,0.197576634199536)
(13,0.215997471053804)
(14,0.236097022515142)
(15,0.25797017346426)
(16,0.281697354093866)
(17,0.307337867817121)
(18,0.334922234009157)
(19,0.364443797619891)
(20,0.395850052545587)
(21,0.42903434886725)
(22,0.463828877111036)
(23,0.5)
(24,0.537247074247846)
(25,0.575205808705361)
(26,0.613456892178211)
(27,0.651540083846052)
(28,0.688973238988476)
};
\end{axis}
\end{tikzpicture}
\end{figure}


%% file: Figure2_random_seeding.tex
\begin{figure}[ht]
\centering
\caption[The probability that team $i$ is allocated to the top groups]{The probability that team $i$ is allocated to the top groups \\
\small{Competition design $D(8+6)/R$}}
\label{Fig2}

\begin{tikzpicture}
\begin{axis}[width = 0.8\textwidth, 
height = 0.4\textwidth,
xmin = 1,
xmax = 28,
ymin = 0,
ymax = 1,
ymajorgrids,
xlabel = Value of $i$,
xlabel style = {font=\small},
tick label style = {/pgf/number format/fixed},
]
\draw (axis cs:\pgfkeysvalueof{/pgfplots/xmin},4/7)  -- (axis cs:\pgfkeysvalueof{/pgfplots/xmax},4/7);
\addplot[red,smooth,very thick] coordinates {
(1,0.887)
(2,0.863996)
(3,0.841968)
(4,0.819196)
(5,0.795257)
(6,0.771647)
(7,0.748318)
(8,0.724963)
(9,0.701051)
(10,0.677177)
(11,0.653839)
(12,0.63035)
(13,0.607591)
(14,0.583799)
(15,0.559989)
(16,0.536216)
(17,0.513483)
(18,0.489525)
(19,0.465883)
(20,0.442073)
(21,0.418053)
(22,0.395178)
(23,0.370918)
(24,0.347603)
(25,0.324238)
(26,0.300594)
(27,0.276911)
(28,0.253184)
};
\end{axis}
\end{tikzpicture}

\end{figure}
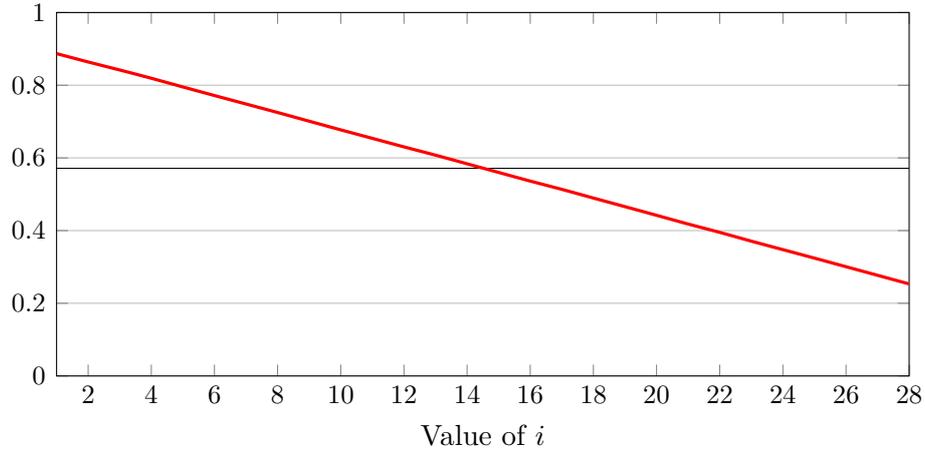

%% file: Figure3_number_of_runs.tex
\begin{figure}[ht]
\centering
\caption[Dependence of some tournament metrics on the number of iterations]{Dependence of some tournament metrics on the number of iterations \\
\small{Competition design $D(8+6)/R$; $\alpha=4$}}
\label{Fig3}

\begin{tikzpicture}
\begin{axis}[
name = axis1,
width = 0.5\textwidth, 
height = 0.35\textwidth,
xmin = 1000,
xmax = 10000000,
title = The proportion of tournament wins for \\ the highest pre-tournament ranked club,
title style = {align=center, font=\small},
ymajorgrids,
xlabel = Number of independent runs,
xlabel style = {font=\small},
xmode = log,
tick label style = {/pgf/number format/precision=3, font=\small},
]
\addplot[red,smooth,very thick] coordinates {
(1000,0.178)
(2500,0.164)
(5000,0.1682)
(10000,0.1698)
(25000,0.16704)
(50000,0.16964)
(100000,0.16535)
(250000,0.166936)
(500000,0.166464)
(1000000,0.166267)
(2500000,0.1667952)
(5000000,0.1669574)
(10000000,0.1668813)
};
\end{axis}

\begin{axis}[
at = {(axis1.south east)},
xshift = 0.1\textwidth,
width = 0.5\textwidth,
height = 0.35\textwidth,
title style = {align=center},
title = The average number of matches between \\ the highest pre-tournament ranked clubs,
title style = {align=center, font=\small},
xmin = 1000,
xmax = 10000000,
ymajorgrids,
xlabel = Number of independent runs,
xlabel style = {font=\small},
xmode = log,
scaled ticks = false,
tick label style = {/pgf/number format/precision=4},
]
\addplot[red,smooth,very thick] coordinates {
(1000,1.061)
(2500,1.0652)
(5000,1.0384)
(10000,1.0277)
(25000,1.04444)
(50000,1.03916)
(100000,1.04108)
(250000,1.043536)
(500000,1.04161)
(1000000,1.039611)
(2500000,1.042144)
(5000000,1.0399222)
(10000000,1.0414988)
};
\end{axis}
\end{tikzpicture}

\end{figure}
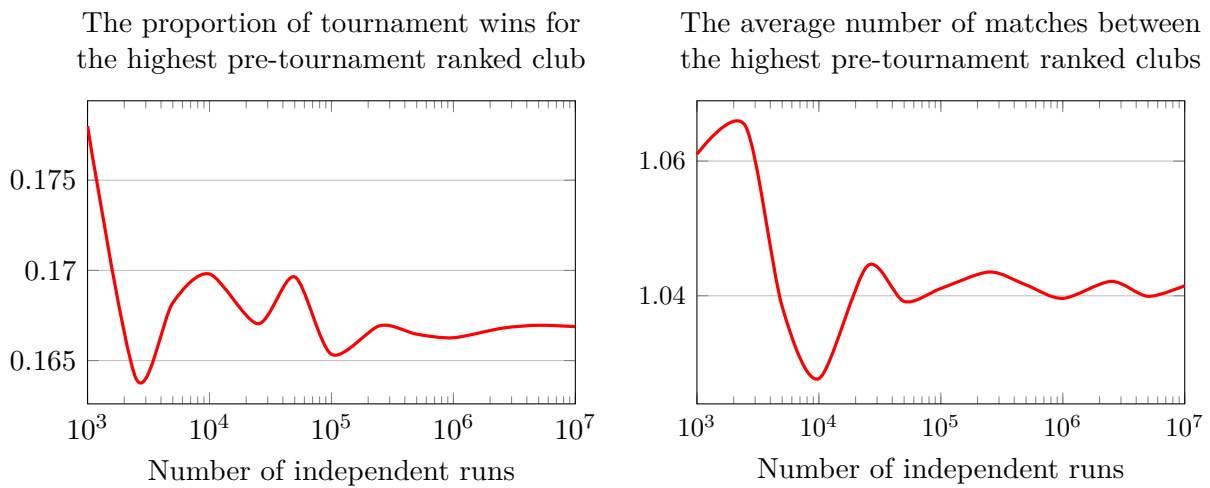

%% file: Figure4_match_number.tex
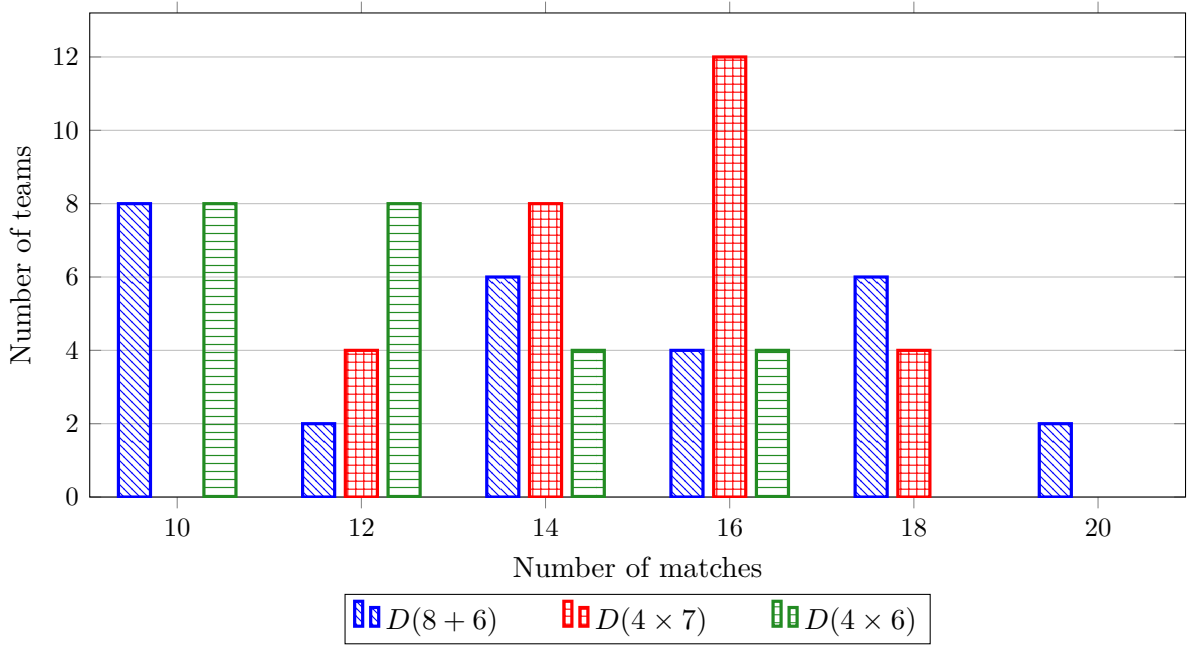
\begin{figure}[ht]
\centering
\caption{The distribution of matches played in a season}
\label{Fig4}

\begin{tikzpicture}
\begin{axis}[width = \textwidth, 
height = 0.5\textwidth,
xlabel = Number of matches,
xlabel style = {font=\small},
ylabel = Number of teams,
ylabel style = {font=\small},
xmin = 9.05,
xmax = 20.95,
xtick distance = 2, 
ybar = 4pt,
ymin = 0,
ymajorgrids = true,
bar width = 12pt,
legend entries={$D(8+6) \qquad$,$D(4 \times 7) \qquad$,$D(4 \times 6)$},
legend style={at={(0.5,-0.2)},anchor = north,legend columns = 4,font=\small}
]
\addplot [blue, pattern color = blue, pattern = north west lines, very thick] coordinates {
(10,8)
(12,2)
(14,6)
(16,4)
(18,6)
(20,2)
};
\addplot [red, pattern color = red, pattern = grid, very thick] coordinates {
(12,4)
(14,8)
(16,12)
(18,4)
};
\addplot [ForestGreen, pattern color = ForestGreen, pattern = horizontal lines, very thick] coordinates {
(10,8)
(12,8)
(14,4)
(16,4)
};
\end{axis}
\end{tikzpicture}
\end{figure}

%% file: Figure5_average_rank.tex
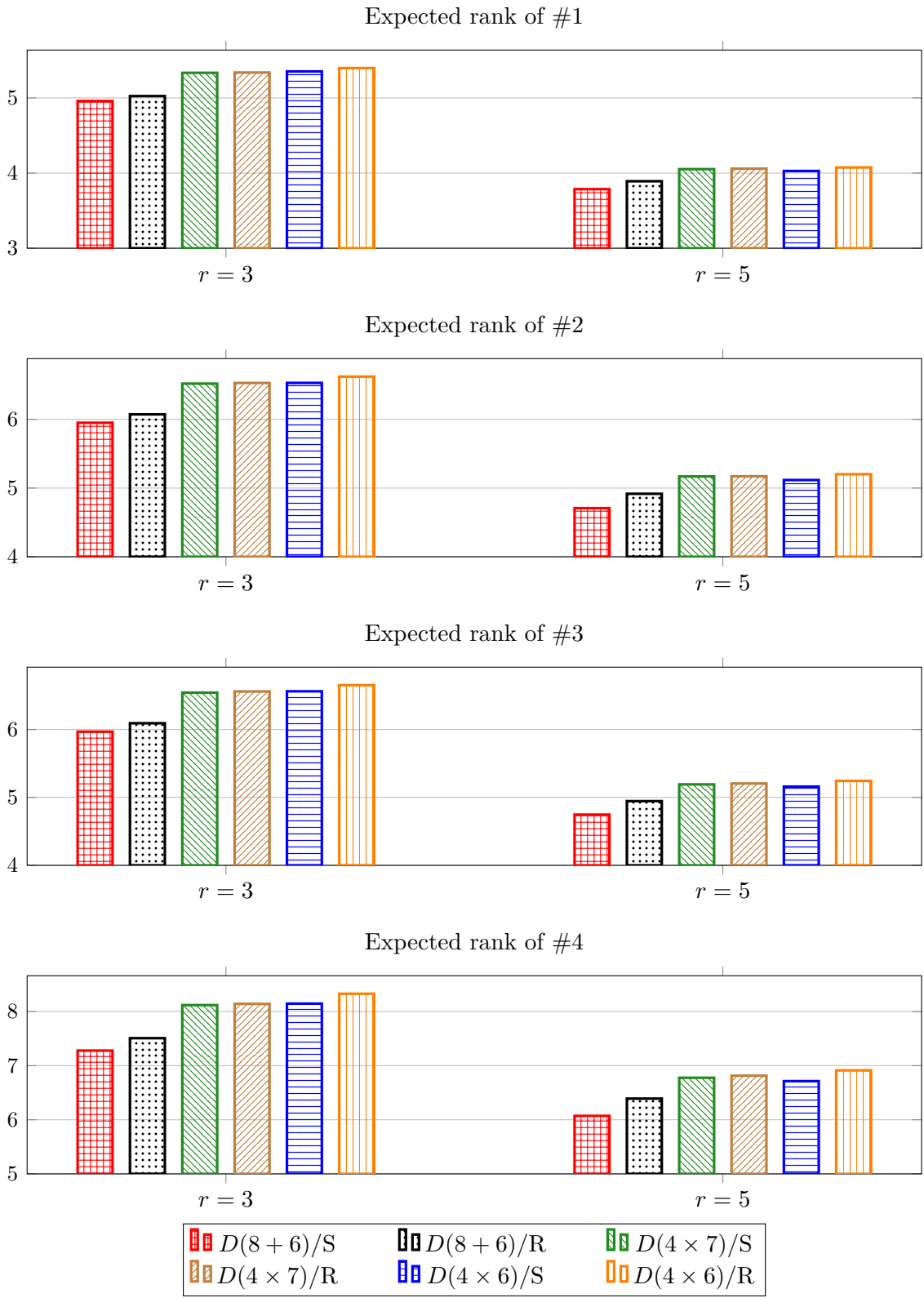
\begin{figure}[ht!]
\centering
\caption{The average pre-tournament ranks of the teams in the Final Four}
\label{Fig5}

\begin{tikzpicture}
\begin{axis}[
name = axis1,
width = \textwidth, 
height = 0.3\textwidth,
title = Expected rank of \#1,
title style = {align=center, font=\small},
symbolic x coords = {$r=3$,$r=5$},
xtick = data,
x tick label style = {font=\small},
enlarge x limits = 0.4,
ybar = 8pt,
ymin = 3,
ymajorgrids = true,
bar width = 16pt,
legend style = {at={(0.5,-0.25)},anchor = north,legend columns = 4,font=\small}
]
\addlegendentry{$D(8+6)$/S $\qquad$}
\addplot [red, pattern color = red, pattern = grid, very thick] coordinates {
($r=3$,4.959461)
($r=5$,3.786206)
};
\addlegendentry{$D(8+6)$/R $\qquad$}
\addplot [black, pattern color = black, pattern = dots, very thick] coordinates {
($r=3$,5.024501)
($r=5$,3.891422)
};
\addlegendentry{$D(4 \times 7)$/S}
\addplot [ForestGreen, pattern color = ForestGreen, pattern = north west lines, very thick] coordinates {
($r=3$,5.334196)
($r=5$,4.052349)
};
\addlegendentry{$D(4 \times 7)$/R $\qquad$}
\addplot [brown, pattern color = brown, pattern = north east lines, very thick] coordinates {
($r=3$,5.337813)
($r=5$,4.058942)
};
\addlegendentry{$D(4 \times 6)$/S $\qquad$}
\addplot [blue, pattern color = blue, pattern = horizontal lines, very thick] coordinates {
($r=3$,5.352392)
($r=5$,4.027914)
};
\addlegendentry{$D(4 \times 6)$/R}
\addplot [orange, pattern color = orange, pattern = vertical lines, very thick] coordinates {
($r=3$,5.397178)
($r=5$,4.074502)
};
\legend{}
\end{axis}
\end{tikzpicture}

\vspace{0.25cm}
\begin{tikzpicture}
\begin{axis}[
name = axis1,
width = \textwidth, 
height = 0.3\textwidth,
title = Expected rank of \#2,
title style = {align=center, font=\small},
symbolic x coords = {$r=3$,$r=5$},
xtick = data,
x tick label style = {font=\small},
enlarge x limits = 0.4,
ybar = 8pt,
ymin = 4,
ymajorgrids = true,
bar width = 16pt,
legend style = {at={(0.5,-0.25)},anchor = north,legend columns = 4,font=\small}
]
\addlegendentry{$D(8+6)$/S $\qquad$}
\addplot [red, pattern color = red, pattern = grid, very thick] coordinates {
($r=3$,5.950545)
($r=5$,4.707498)
};
\addlegendentry{$D(8+6)$/R $\qquad$}
\addplot [black, pattern color = black, pattern = dots, very thick] coordinates {
($r=3$,6.072698)
($r=5$,4.916781)
};
\addlegendentry{$D(4 \times 7)$/S}
\addplot [ForestGreen, pattern color = ForestGreen, pattern = north west lines, very thick] coordinates {
($r=3$,6.520078)
($r=5$,5.169311)
};
\addlegendentry{$D(4 \times 7)$/R $\qquad$}
\addplot [brown, pattern color = brown, pattern = north east lines, very thick] coordinates {
($r=3$,6.530128)
($r=5$,5.171564)
};
\addlegendentry{$D(4 \times 6)$/S $\qquad$}
\addplot [blue, pattern color = blue, pattern = horizontal lines, very thick] coordinates {
($r=3$,6.530505)
($r=5$,5.117058)
};
\addlegendentry{$D(4 \times 6)$/R}
\addplot [orange, pattern color = orange, pattern = vertical lines, very thick] coordinates {
($r=3$,6.621965)
($r=5$,5.201975)
};
\legend{}
\end{axis}
\end{tikzpicture}

\vspace{0.25cm}
\begin{tikzpicture}
\begin{axis}[
name = axis1,
width = \textwidth, 
height = 0.3\textwidth,
title = Expected rank of \#3,
title style = {align=center, font=\small},
symbolic x coords = {$r=3$,$r=5$},
xtick = data,
x tick label style = {font=\small},
enlarge x limits = 0.4,
ybar = 8pt,
ymin = 4,
ymajorgrids = true,
bar width = 16pt,
legend style = {at={(0.5,-0.25)},anchor = north,legend columns = 4,font=\small}
]
\addlegendentry{$D(8+6)$/S $\qquad$}
\addplot [red, pattern color = red, pattern = grid, very thick] coordinates {
($r=3$,5.967933)
($r=5$,4.747038)
};
\addlegendentry{$D(8+6)$/R $\qquad$}
\addplot [black, pattern color = black, pattern = dots, very thick] coordinates {
($r=3$,6.092494)
($r=5$,4.945628)
};
\addlegendentry{$D(4 \times 7)$/S}
\addplot [ForestGreen, pattern color = ForestGreen, pattern = north west lines, very thick] coordinates {
($r=3$,6.5433)
($r=5$,5.192066)
};
\addlegendentry{$D(4 \times 7)$/R $\qquad$}
\addplot [brown, pattern color = brown, pattern = north east lines, very thick] coordinates {
($r=3$,6.56065)
($r=5$,5.20531)
};
\addlegendentry{$D(4 \times 6)$/S $\qquad$}
\addplot [blue, pattern color = blue, pattern = horizontal lines, very thick] coordinates {
($r=3$,6.562582)
($r=5$,5.159418)
};
\addlegendentry{$D(4 \times 6)$/R}
\addplot [orange, pattern color = orange, pattern = vertical lines, very thick] coordinates {
($r=3$,6.65301)
($r=5$,5.24388)
};
\legend{}
\end{axis}
\end{tikzpicture}

\vspace{0.25cm}
\begin{tikzpicture}
\begin{axis}[
name = axis1,
width = \textwidth, 
height = 0.3\textwidth,
title = Expected rank of \#4,
title style = {align=center, font=\small},
symbolic x coords = {$r=3$,$r=5$},
xtick = data,
x tick label style = {font=\small},
enlarge x limits = 0.4,
ybar = 8pt,
ymin = 5,
ymajorgrids = true,
bar width = 16pt,
legend style = {at={(0.5,-0.25)},anchor = north,legend columns = 3,font=\small}
]
\addlegendentry{$D(8+6)$/S $\qquad$}
\addplot [red, pattern color = red, pattern = grid, very thick] coordinates {
($r=3$,7.278712)
($r=5$,6.071874)
};
\addlegendentry{$D(8+6)$/R $\qquad$}
\addplot [black, pattern color = black, pattern = dots, very thick] coordinates {
($r=3$,7.508157)
($r=5$,6.391243)
};
\addlegendentry{$D(4 \times 7)$/S}
\addplot [ForestGreen, pattern color = ForestGreen, pattern = north west lines, very thick] coordinates {
($r=3$,8.119443)
($r=5$,6.776301)
};
\addlegendentry{$D(4 \times 7)$/R $\qquad$}
\addplot [brown, pattern color = brown, pattern = north east lines, very thick] coordinates {
($r=3$,8.141913)
($r=5$,6.814659)
};
\addlegendentry{$D(4 \times 6)$/S $\qquad$}
\addplot [blue, pattern color = blue, pattern = horizontal lines, very thick] coordinates {
($r=3$,8.146549)
($r=5$,6.713871)
};
\addlegendentry{$D(4 \times 6)$/R}
\addplot [orange, pattern color = orange, pattern = vertical lines, very thick] coordinates {
($r=3$,8.329356)
($r=5$,6.912722)
};
\end{axis}
\end{tikzpicture}

\end{figure}

%% file: Figure6_quality_competitive_balance.tex
\begin{figure}[ht!]
\centering
\caption{Characteristics of all matches played}
\label{Fig6}

\begin{tikzpicture}
\begin{axis}[
name = axis1,
width = \textwidth, 
height = 0.35\textwidth,
title = Expected quality,
title style = {align=center, font=\small},
symbolic x coords = {$r=3$,$r=5$},
xtick = data,
x tick label style = {font=\small},
enlarge x limits = 0.4,
ybar = 8pt,
ymajorgrids = true,
bar width = 16pt,
legend style = {at={(0.5,-0.25)},anchor = north,legend columns = 4,font=\small}
]
\addlegendentry{$D(8+6)$/S $\qquad$}
\addplot [red, pattern color = red, pattern = grid, very thick] coordinates {
($r=3$,48.7995269622642)
($r=5$,48.4065636792453)
};
\addlegendentry{$D(8+6)$/R $\qquad$}
\addplot [black, pattern color = black, pattern = dots, very thick] coordinates {
($r=3$,50.335464754717)
($r=5$,49.8892061886792)
};
\addlegendentry{$D(4 \times 7)$/S}
\addplot [ForestGreen, pattern color = ForestGreen, pattern = north west lines, very thick] coordinates {
($r=3$,55.5171513962264)
($r=5$,55.0402074716981)
};
\addlegendentry{$D(4 \times 7)$/R $\qquad$}
\addplot [brown, pattern color = brown, pattern = north east lines, very thick] coordinates {
($r=3$,55.5653351132075)
($r=5$,55.0854902075472)
};
\addlegendentry{$D(4 \times 6)$/S $\qquad$}
\addplot [blue, pattern color = blue, pattern = horizontal lines, very thick] coordinates {
($r=3$,47.2677006486486)
($r=5$,46.5209995135135)
};
\addlegendentry{$D(4 \times 6)$/R}
\addplot [orange, pattern color = orange, pattern = vertical lines, very thick] coordinates {
($r=3$,47.4394877027027)
($r=5$,46.7245787297297)
};
\legend{}
\end{axis}
\end{tikzpicture}

\vspace{0.25cm}
\begin{tikzpicture}
\begin{axis}[
name = axis1,
width = \textwidth, 
height = 0.35\textwidth,
title = Expected competitive balance,
title style = {align=center, font=\small},
symbolic x coords = {$r=3$,$r=5$},
xtick = data,
x tick label style = {font=\small},
enlarge x limits = 0.4,
ybar = 8pt,
ymajorgrids = true,
bar width = 16pt,
legend style = {at={(0.5,-0.25)},anchor = north,legend columns = 3,font=\small}
]
\addlegendentry{$D(8+6)$/S $\qquad$}
\addplot [red, pattern color = red, pattern = grid, very thick] coordinates {
($r=3$,10.6696628490566)
($r=5$,10.5445243018868)
};
\addlegendentry{$D(8+6)$/R $\qquad$}
\addplot [black, pattern color = black, pattern = dots, very thick] coordinates {
($r=3$,16.4266165471698)
($r=5$,16.2365556037736)
};
\addlegendentry{$D(4 \times 7)$/S}
\addplot [ForestGreen, pattern color = ForestGreen, pattern = north west lines, very thick] coordinates {
($r=3$,20.2446589811321)
($r=5$,20.0272408301887)
};
\addlegendentry{$D(4 \times 7)$/R $\qquad$}
\addplot [brown, pattern color = brown, pattern = north east lines, very thick] coordinates {
($r=3$,19.0181745849057)
($r=5$,18.7982540377358)
};
\addlegendentry{$D(4 \times 6)$/S $\qquad$}
\addplot [blue, pattern color = blue, pattern = horizontal lines, very thick] coordinates {
($r=3$,17.8198014054054)
($r=5$,17.6198604054054)
};
\addlegendentry{$D(4 \times 6)$/R}
\addplot [orange, pattern color = orange, pattern = vertical lines, very thick] coordinates {
($r=3$,16.484567)
($r=5$,16.2881449189189)
};
\end{axis}
\end{tikzpicture}

\end{figure}
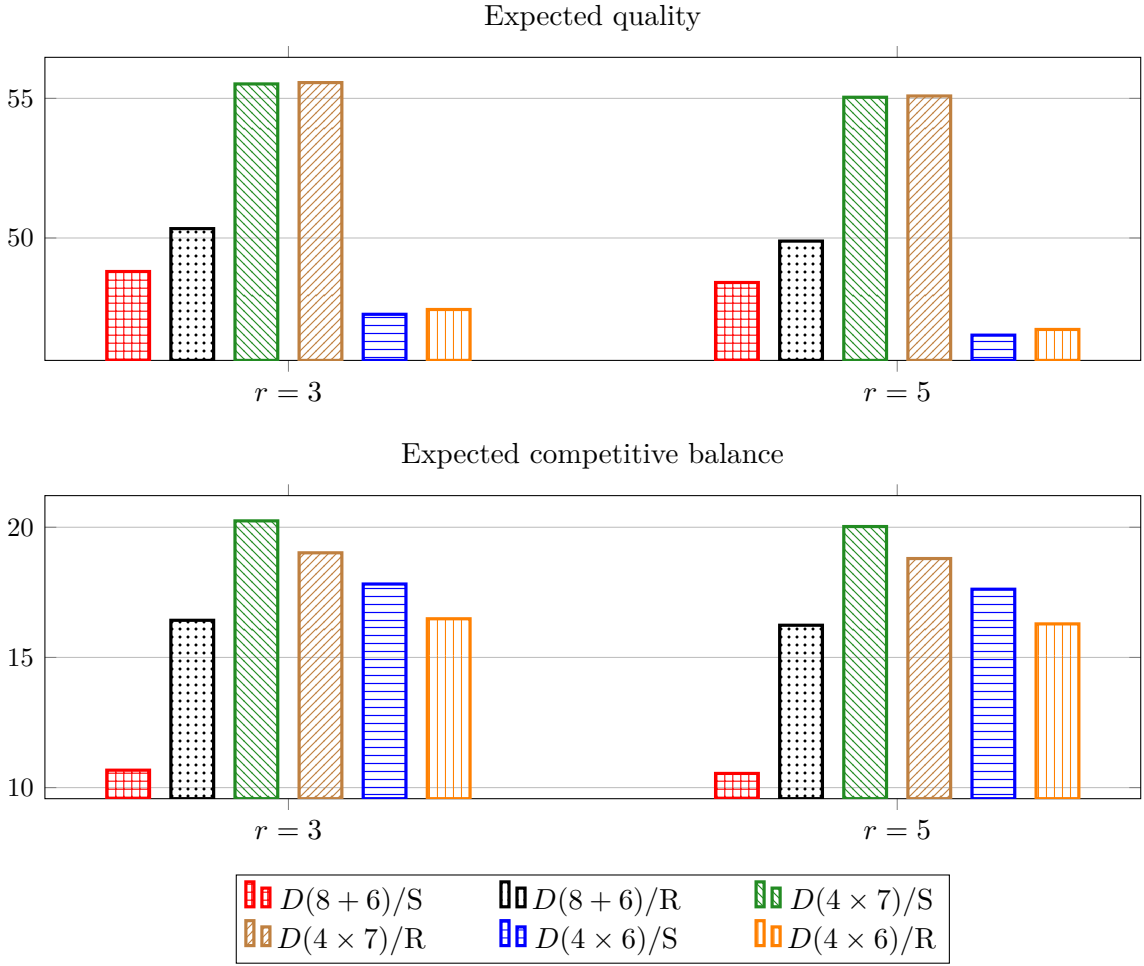

%% file: Figure7_matches_points.tex
\begin{figure}[ht]
\centering
\caption[Individual team statistics]{Individual team statistics ($r = 3$)}
\label{Fig7}

\begin{subfigure}{\textwidth}
\caption{Correct team identification}
\label{Fig7a}

\begin{tikzpicture}
\begin{axis}[
name = axis1,
width = 0.5\textwidth, 
height = 0.35\textwidth,
title = Expected number of matches played,
title style = {align=center, font=\small},
xmin = 1,
xmax = 28,
ymajorgrids,
xlabel = Value of $i$,
xlabel style = {font=\small},
scaled ticks = false,
y tick label style = {/pgf/number format/1000 sep=\,},
]
\draw (axis cs:\pgfkeysvalueof{/pgfplots/xmin},1.0)  -- (axis cs:\pgfkeysvalueof{/pgfplots/xmax},1.0);
\addplot[red,very thick] coordinates {
(1,17.824972)
(2,17.730596)
(3,17.648802)
(4,17.528888)
(5,17.402178)
(6,17.25336)
(7,17.084956)
(8,16.908868)
(9,16.707074)
(10,16.516498)
(11,16.284474)
(12,16.084292)
(13,15.826668)
(14,15.633038)
(15,15.38605)
(16,15.202966)
(17,12.277392)
(18,12.039192)
(19,11.756506)
(20,11.538442)
(21,11.265338)
(22,11.076348)
(23,10.851896)
(24,10.697692)
(25,10.529712)
(26,10.417406)
(27,10.299724)
(28,10.226672)
};
\addplot[black,very thick, loosely dotted] coordinates {
(1,17.502256)
(2,17.361316)
(3,17.18448)
(4,16.996074)
(5,16.789594)
(6,16.566626)
(7,16.330868)
(8,16.082202)
(9,15.815006)
(10,15.55535)
(11,15.268856)
(12,14.984752)
(13,14.693272)
(14,14.391686)
(15,14.105086)
(16,13.809954)
(17,13.523534)
(18,13.248564)
(19,12.972846)
(20,12.715452)
(21,12.471472)
(22,12.239586)
(23,12.025742)
(24,11.817342)
(25,11.631674)
(26,11.457948)
(27,11.302322)
(28,11.15614)
};
\addplot[ForestGreen,very thick,densely dashed] coordinates {
(1,16.687406)
(2,16.613302)
(3,16.538858)
(4,16.454002)
(5,16.42212)
(6,16.330908)
(7,16.235748)
(8,16.129684)
(9,16.083964)
(10,15.966988)
(11,15.845406)
(12,15.725982)
(13,15.606698)
(14,15.470236)
(15,15.322916)
(16,15.177212)
(17,14.996984)
(18,14.840484)
(19,14.676618)
(20,14.513228)
(21,14.257044)
(22,14.088168)
(23,13.923258)
(24,13.75639)
(25,13.310356)
(26,13.152124)
(27,13.006044)
(28,12.867872)
};
\addplot[brown,very thick, loosely dashed] coordinates {
(1,16.748536)
(2,16.674798)
(3,16.601744)
(4,16.513044)
(5,16.428656)
(6,16.33911)
(7,16.236856)
(8,16.13179)
(9,16.026766)
(10,15.907136)
(11,15.790638)
(12,15.66028)
(13,15.526612)
(14,15.380812)
(15,15.23359)
(16,15.076754)
(17,14.91281)
(18,14.753156)
(19,14.584254)
(20,14.412916)
(21,14.236292)
(22,14.06141)
(23,13.88617)
(24,13.71209)
(25,13.53956)
(26,13.370704)
(27,13.206838)
(28,13.046678)
};
\addplot[blue,smooth,very thick,loosely dashdotdotted] coordinates {
(1,14.20344)
(2,14.082284)
(3,13.960114)
(4,13.830086)
(5,13.67058)
(6,13.521732)
(7,13.373014)
(8,13.21664)
(9,13.019102)
(10,12.85289)
(11,12.68572)
(12,12.518268)
(13,12.253608)
(14,12.082226)
(15,11.907826)
(16,11.741642)
(17,11.416674)
(18,11.258492)
(19,11.113426)
(20,10.974868)
(21,10.737286)
(22,10.62565)
(23,10.523482)
(24,10.43095)
};
\addplot[orange,smooth,very thick, densely dashdotdotted] coordinates {
(1,14.14687)
(2,14.024124)
(3,13.892198)
(4,13.750604)
(5,13.604386)
(6,13.45139)
(7,13.293628)
(8,13.12863)
(9,12.959638)
(10,12.782376)
(11,12.605128)
(12,12.42793)
(13,12.244114)
(14,12.05757)
(15,11.876102)
(16,11.70202)
(17,11.526068)
(18,11.363494)
(19,11.203208)
(20,11.052998)
(21,10.911824)
(22,10.78101)
(23,10.66217)
(24,10.55252)
};
\end{axis}

\begin{axis}[
at = {(axis1.south east)},
xshift = 0.1\textwidth,
width = 0.5\textwidth, 
height = 0.35\textwidth,
title = Expected winning percentage,
title style = {align=center, font=\small},
xmin = 1,
xmax = 28,
ymajorgrids,
xlabel = Value of $i$,
xlabel style = {font=\small},
scaled ticks = false,
y tick label style={/pgf/number format/fixed},
]
\draw (axis cs:\pgfkeysvalueof{/pgfplots/xmin},0.5)  -- (axis cs:\pgfkeysvalueof{/pgfplots/xmax},0.5);
\addplot[red,very thick] coordinates {
(1,0.620600413846372)
(2,0.607380315923954)
(3,0.591495332091096)
(4,0.57786751789389)
(5,0.560903181199503)
(6,0.546274754598525)
(7,0.528341425052543)
(8,0.513210582754564)
(9,0.494179591231834)
(10,0.478571486522143)
(11,0.459349684859333)
(12,0.442841562438682)
(13,0.422198974540946)
(14,0.405510304523024)
(15,0.384480552188508)
(16,0.367368577947224)
(17,0.600580644488667)
(18,0.585759742015909)
(19,0.564783873712139)
(20,0.548334688513406)
(21,0.525176075498134)
(22,0.506970167423414)
(23,0.480938630447619)
(24,0.460764994916661)
(25,0.431571822667135)
(26,0.410009075195879)
(27,0.37843984945616)
(28,0.355740459848522)
};
\addplot[black,very thick, loosely dotted] coordinates {
(1,0.660334244911056)
(2,0.64934507268919)
(3,0.638009704105099)
(4,0.626267630983485)
(5,0.614648454274713)
(6,0.60269001062739)
(7,0.590482943098922)
(8,0.578224486920386)
(9,0.565462700425153)
(10,0.552761397204177)
(11,0.539483049679688)
(12,0.526457561660013)
(13,0.512964641231715)
(14,0.499025895923521)
(15,0.485099062848677)
(16,0.471203234999914)
(17,0.456369984354681)
(18,0.441752328780689)
(19,0.426613944233979)
(20,0.411311921904152)
(21,0.39538628639827)
(22,0.379205309722077)
(23,0.362832164535045)
(24,0.345884294454709)
(25,0.32869404696177)
(26,0.311083625095872)
(27,0.293012798608994)
(28,0.274970195784563)
};
\addplot[ForestGreen,very thick,densely dashed] coordinates {
(1,0.699338770807158)
(2,0.688436711738582)
(3,0.677575984992434)
(4,0.666297901264385)
(5,0.648516695773749)
(6,0.636543969263681)
(7,0.62439531581791)
(8,0.611510430086541)
(9,0.59160484318418)
(10,0.578094628742754)
(11,0.564238429737932)
(12,0.550084694234039)
(13,0.526884226246961)
(14,0.511873509880521)
(15,0.496033065768944)
(16,0.480002980784613)
(17,0.453754768292078)
(18,0.437049357689412)
(19,0.419835823212132)
(20,0.402226644547994)
(21,0.373222457614636)
(22,0.355089604269341)
(23,0.337093372829836)
(24,0.318730131960493)
(25,0.28707909841029)
(26,0.268920746185179)
(27,0.250898582228386)
(28,0.233061845812579)
};
\addplot[brown,very thick, loosely dashed] coordinates {
(1,0.687166627578673)
(2,0.676096765909848)
(3,0.664697817289557)
(4,0.652704734511699)
(5,0.640717779957167)
(6,0.628241011903341)
(7,0.615300770050557)
(8,0.601426810044019)
(9,0.587532194579992)
(10,0.573049101987938)
(11,0.558327788908846)
(12,0.54321780964325)
(13,0.527433415609278)
(14,0.510905861146993)
(15,0.494139267237729)
(16,0.476996639992932)
(17,0.459282589934426)
(18,0.441851221528465)
(19,0.423431325318388)
(20,0.404786928613197)
(21,0.386162351825883)
(22,0.36675852563861)
(23,0.348036643653362)
(24,0.328473267022022)
(25,0.309217507806753)
(26,0.290042020225711)
(27,0.271009003063413)
(28,0.251837594213638)
};
\addplot[blue,smooth,very thick,loosely dashdotdotted] coordinates {
(1,0.67426186895569)
(2,0.662638532215371)
(3,0.650680646304178)
(4,0.638439847734859)
(5,0.617447028582547)
(6,0.604174006702692)
(7,0.590675744450727)
(8,0.577289462374703)
(9,0.553855096918359)
(10,0.539558184968517)
(11,0.524830045121601)
(12,0.509872292237233)
(13,0.483916818621911)
(14,0.468095779701522)
(15,0.452046662421839)
(16,0.435911859687086)
(17,0.407552497338542)
(18,0.390672747291556)
(19,0.373790674450885)
(20,0.356726021670602)
(21,0.324714830172168)
(22,0.307560384541181)
(23,0.289968662463622)
(24,0.272264750574013)
};
\addplot[orange,smooth,very thick, densely dashdotdotted] coordinates {
(1,0.661512546591578)
(2,0.6494467675842)
(3,0.637175701066167)
(4,0.623963209179757)
(5,0.61076479306012)
(6,0.596687777248299)
(7,0.582619131511729)
(8,0.568006410417538)
(9,0.553036589447946)
(10,0.537693931081358)
(11,0.52165126764282)
(12,0.505489329276879)
(13,0.488737608944183)
(14,0.471821187851283)
(15,0.454875766476239)
(16,0.437595902245937)
(17,0.41931194575635)
(18,0.401637295712041)
(19,0.383570223814465)
(20,0.365595560589082)
(21,0.347513303000488)
(22,0.328999787589474)
(23,0.310587432014309)
(24,0.292338796799248)
};
\end{axis}
\end{tikzpicture}
\end{subfigure}

\vspace{0.25cm}
\begin{subfigure}{\textwidth}
\caption{Erroneous team identification}
\label{Fig7b}

\begin{tikzpicture}
\begin{axis}[
name = axis1,
width = 0.5\textwidth, 
height = 0.35\textwidth,
title = Expected number of matches played,
title style = {align=center, font=\small},
xmin = 1,
xmax = 28,
ymajorgrids,
xlabel = Value of $i$,
xlabel style = {font=\small},
scaled ticks = false,
y tick label style={/pgf/number format/1000 sep=\,},
]
\draw (axis cs:\pgfkeysvalueof{/pgfplots/xmin},1.0)  -- (axis cs:\pgfkeysvalueof{/pgfplots/xmax},1.0);
\addplot[red,very thick] coordinates {
(1,17.824986)
(2,17.744746)
(3,17.66911)
(4,17.55864)
(5,17.441976)
(6,17.306808)
(7,17.153188)
(8,16.98931)
(9,13.955628)
(10,16.601816)
(11,16.410842)
(12,16.176928)
(13,15.979226)
(14,15.723828)
(15,15.531514)
(16,15.288322)
(17,15.116364)
(18,11.902922)
(19,11.581314)
(20,11.379904)
(21,11.114352)
(22,10.943306)
(23,10.740324)
(24,10.604186)
(25,10.454836)
(26,10.356086)
(27,10.257964)
(28,10.191574)
};
\addplot[black,very thick, loosely dotted] coordinates {
(1,17.503756)
(2,17.35275)
(3,17.183938)
(4,16.992412)
(5,16.786234)
(6,16.56632)
(7,16.330322)
(8,16.078848)
(9,14.964168)
(10,15.656316)
(11,15.380696)
(12,15.094648)
(13,14.808182)
(14,14.50501)
(15,14.212432)
(16,13.916124)
(17,13.633498)
(18,13.242492)
(19,12.975168)
(20,12.71612)
(21,12.473496)
(22,12.239744)
(23,12.02294)
(24,11.813532)
(25,11.633146)
(26,11.459954)
(27,11.302434)
(28,11.15532)
};
\addplot[ForestGreen,very thick,densely dashed] coordinates {
(1,16.682446)
(2,16.60491)
(3,16.530868)
(4,16.450128)
(5,16.41099)
(6,16.32688)
(7,16.228046)
(8,16.130644)
(9,16.116628)
(10,15.971828)
(11,15.854848)
(12,15.725618)
(13,15.597588)
(14,15.461962)
(15,15.321224)
(16,15.169284)
(17,15.014002)
(18,14.844648)
(19,14.682168)
(20,14.520652)
(21,14.256872)
(22,14.08698)
(23,13.91989)
(24,13.754968)
(25,13.31049)
(26,13.154584)
(27,13.004012)
(28,12.866842)
};
\addplot[brown,very thick, loosely dashed] coordinates {
(1,16.749852)
(2,16.674454)
(3,16.600304)
(4,16.513944)
(5,16.426178)
(6,16.332142)
(7,16.237244)
(8,16.133158)
(9,16.035344)
(10,15.908672)
(11,15.785826)
(12,15.655542)
(13,15.525096)
(14,15.383718)
(15,15.233992)
(16,15.079438)
(17,14.920406)
(18,14.748562)
(19,14.582144)
(20,14.408954)
(21,14.236532)
(22,14.062194)
(23,13.884234)
(24,13.712992)
(25,13.541906)
(26,13.371196)
(27,13.205984)
(28,13.049992)
};
\addplot[blue,smooth,very thick,loosely dashdotted] coordinates {
(1,14.197314)
(2,14.084502)
(3,13.958532)
(4,13.826554)
(5,13.661528)
(6,13.519964)
(7,13.367814)
(8,13.214908)
(9,12.868386)
(10,12.829644)
(11,12.66395)
(12,12.491044)
(13,12.319798)
(14,12.03755)
(15,11.868284)
(16,11.699844)
(17,11.537228)
(18,11.320388)
(19,11.170082)
(20,11.029782)
(21,10.74068)
(22,10.6281)
(23,10.527458)
(24,10.436666)
};
\addplot[orange,smooth,very thick, densely dashdotted] coordinates {
(1,14.151968)
(2,14.022572)
(3,13.887556)
(4,13.750824)
(5,13.605816)
(6,13.454082)
(7,13.290106)
(8,13.127804)
(9,12.930426)
(10,12.786202)
(11,12.604442)
(12,12.427736)
(13,12.244814)
(14,12.063492)
(15,11.881198)
(16,11.706006)
(17,11.530244)
(18,11.36173)
(19,11.20477)
(20,11.055244)
(21,10.914802)
(22,10.782662)
(23,10.662668)
(24,10.552836)
};
\end{axis}

\begin{axis}[
at = {(axis1.south east)},
xshift = 0.1\textwidth,
width = 0.5\textwidth, 
height = 0.35\textwidth,
title = Expected winning percentage,
title style = {align=center, font=\small},
xmin = 1,
xmax = 28,
ymajorgrids,
xlabel = Value of $i$,
xlabel style = {font=\small},
scaled ticks = false,
y tick label style={/pgf/number format/fixed},
legend entries={$D(8+6)/S \qquad$,$D(8+6)/R \qquad$,$D(4 \times 7)/S$,$D(4 \times 7)/R \qquad$,$D(4 \times 6)/S \qquad$,$D(4 \times 6)/R$},
legend style = {at={(-0.1,-0.3)},anchor=north,legend columns = 3,font=\small}
]
\draw (axis cs:\pgfkeysvalueof{/pgfplots/xmin},0.5)  -- (axis cs:\pgfkeysvalueof{/pgfplots/xmax},0.5);
\addplot[red,very thick] coordinates {
(1,0.626517967531643)
(2,0.614102563091069)
(3,0.598044723248653)
(4,0.584627909678654)
(5,0.567688488964782)
(6,0.553556149695542)
(7,0.53569214072626)
(8,0.520782656858931)
(9,0.695539032711391)
(10,0.484779375942969)
(11,0.468877709017002)
(12,0.449115926089304)
(13,0.432612255437153)
(14,0.411872795861161)
(15,0.395240283722501)
(16,0.373698761708446)
(17,0.357067810751316)
(18,0.58272859386964)
(19,0.553197590532473)
(20,0.536866128220414)
(21,0.513588736437356)
(22,0.495415370821213)
(23,0.469462653081974)
(24,0.449511824858598)
(25,0.420591198178527)
(26,0.399395582462332)
(27,0.368150736344951)
(28,0.346396837230442)
};
\addplot[black,very thick, loosely dotted] coordinates {
(1,0.661066573368596)
(2,0.649777009407731)
(3,0.638239907522944)
(4,0.626907704450669)
(5,0.61506910960493)
(6,0.603028312866104)
(7,0.5909086789593)
(8,0.578297960152369)
(9,0.579775233745037)
(10,0.551427232306757)
(11,0.537953939145537)
(12,0.524938243011695)
(13,0.511168420269281)
(14,0.497394624340142)
(15,0.483176841233084)
(16,0.469187541013575)
(17,0.45460490037113)
(18,0.441672081055439)
(19,0.426654591293153)
(20,0.41146709845456)
(21,0.395498583556687)
(22,0.379448295650628)
(23,0.362490206222438)
(24,0.345648871142009)
(25,0.328337407610976)
(26,0.310668437238055)
(27,0.292797905300752)
(28,0.274285542682774)
};
\addplot[ForestGreen,very thick,densely dashed] coordinates {
(1,0.699432325451556)
(2,0.688594759020073)
(3,0.677508646248945)
(4,0.666427580381137)
(5,0.648131709299683)
(6,0.636272515018179)
(7,0.623811024444964)
(8,0.611345833433557)
(9,0.579552496961523)
(10,0.576255391680902)
(11,0.562056602497861)
(12,0.548282236030406)
(13,0.533125762778194)
(14,0.509448930219852)
(15,0.494107259315574)
(16,0.477688202027202)
(17,0.46133875564956)
(18,0.441357855033006)
(19,0.423961434033448)
(20,0.406637181305633)
(21,0.373487325971644)
(22,0.355249599275359)
(23,0.336883624798759)
(24,0.318710301616114)
(25,0.287600306224639)
(26,0.269109308207694)
(27,0.250739848594418)
(28,0.233163118036267)
};
\addplot[brown,very thick, loosely dashed] coordinates {
(1,0.687175564297523)
(2,0.676068133925105)
(3,0.664571925911718)
(4,0.652493068887723)
(5,0.640535613336225)
(6,0.627777728114291)
(7,0.614747921506876)
(8,0.601367382628993)
(9,0.584355907799671)
(10,0.573537250626576)
(11,0.559154649240401)
(12,0.543610754581349)
(13,0.527597252860788)
(14,0.511766076315232)
(15,0.494839041532909)
(16,0.477569853730623)
(17,0.460171928297394)
(18,0.441444799838791)
(19,0.423041700863741)
(20,0.404773656713735)
(21,0.386158511075591)
(22,0.367115615102451)
(23,0.347951424615863)
(24,0.328809059321263)
(25,0.309422248241865)
(26,0.2903490458146)
(27,0.270866903973229)
(28,0.251978545274204)
};
\addplot[blue,smooth,very thick,loosely dashdotted] coordinates {
(1,0.674147307018778)
(2,0.663025359363079)
(3,0.651099628528272)
(4,0.638618559620857)
(5,0.617400776838433)
(6,0.604283783595873)
(7,0.59082397466033)
(8,0.577394031044333)
(9,0.539339587730738)
(10,0.536995726459752)
(11,0.522480821544621)
(12,0.507470392386737)
(13,0.491937611314731)
(14,0.465412895481223)
(15,0.449671072920062)
(16,0.433480566065667)
(17,0.416767181856855)
(18,0.396361679476004)
(19,0.379437321946249)
(20,0.362339980971519)
(21,0.324755508962189)
(22,0.307237323698497)
(23,0.290013885593274)
(24,0.272679321154859)
};
\addplot[orange,smooth,very thick, densely dashdotted] coordinates {
(1,0.661780396903102)
(2,0.649116866720314)
(3,0.636791455602411)
(4,0.624038748514271)
(5,0.610696925491275)
(6,0.596630598802653)
(7,0.582554646291008)
(8,0.568042454015919)
(9,0.549595968454558)
(10,0.537903436845437)
(11,0.521844362487447)
(12,0.506089765666088)
(13,0.489406862366386)
(14,0.472208544590571)
(15,0.455081634023774)
(16,0.437819440721284)
(17,0.4199037765376)
(18,0.40165036486521)
(19,0.383956832670372)
(20,0.366106528268395)
(21,0.347659810961298)
(22,0.329351045224268)
(23,0.310836743674285)
(24,0.292316018177483)
};
\end{axis}
\end{tikzpicture}
\end{subfigure}

\end{figure}


%% file: Figure8_expected_prize.tex
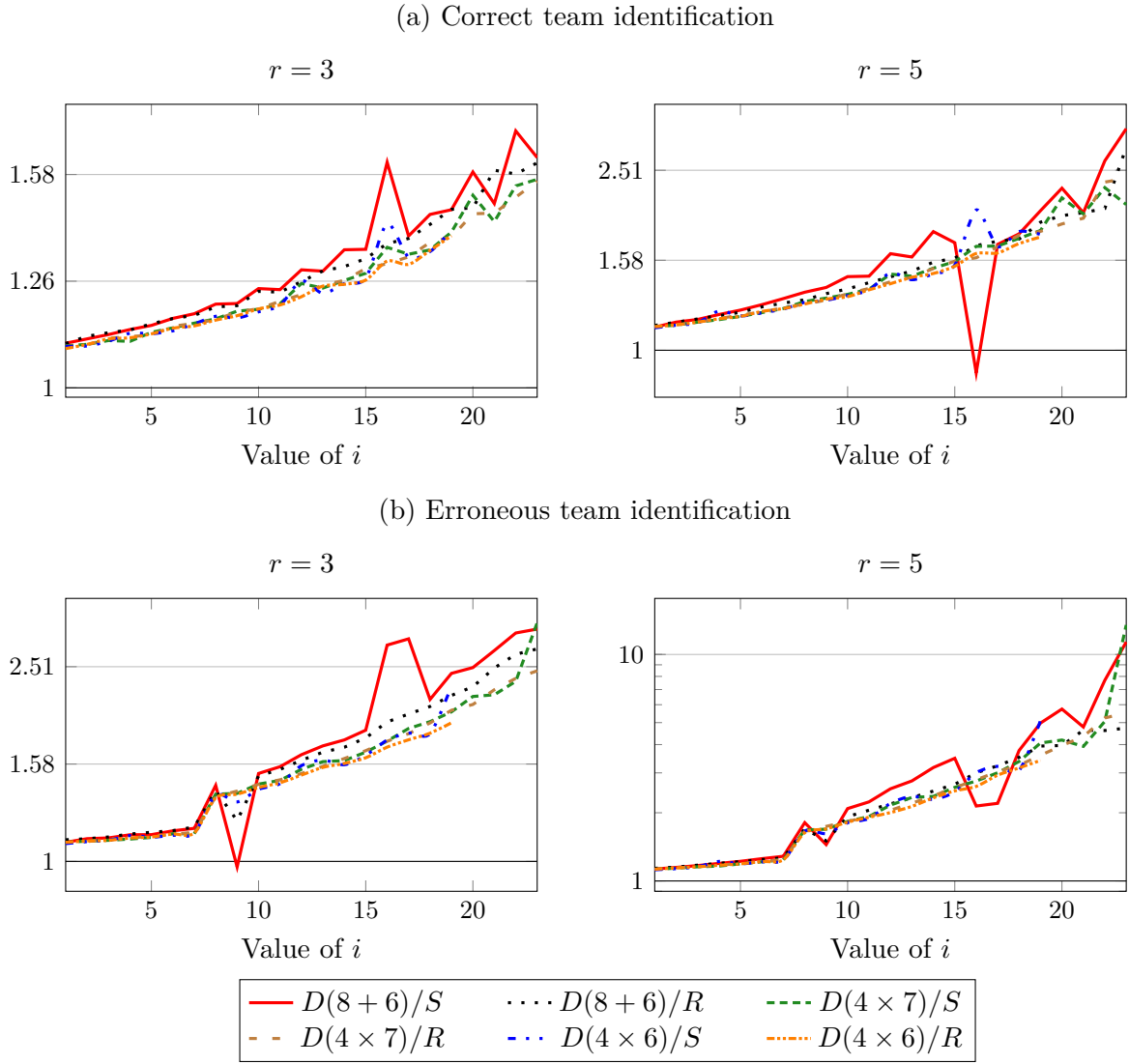
\begin{figure}[ht]
\centering
\caption{The ratio between the expected prizes of team $i$ and team $i+1$}
\label{Fig8}

\begin{subfigure}{\textwidth}
\caption{Correct team identification}
\label{Fig8a}

\begin{tikzpicture}
\begin{axis}[
name = axis1,
width = 0.5\textwidth, 
height = 0.35\textwidth,
title = {$r=3$},
title style = {align=center, font=\small},
xmin = 1,
xmax = 23,
ymin = 0.98,
ymajorgrids,
ymode = log,
xlabel = Value of $i$,
xlabel style = {font=\small},
log ticks with fixed point,
y tick label style={/pgf/number format/1000 sep=\,},
]
\draw (axis cs:\pgfkeysvalueof{/pgfplots/xmin},1.0)  -- (axis cs:\pgfkeysvalueof{/pgfplots/xmax},1.0);
\addplot[red,very thick] coordinates {
(1,1.10108537143395)
(2,1.11151144965598)
(3,1.12193754021951)
(4,1.1341503368451)
(5,1.14429162855241)
(6,1.16173492839253)
(7,1.17351934239475)
(8,1.19807932895213)
(9,1.19962389700564)
(10,1.23903113876761)
(11,1.23612351576834)
(12,1.29023181026397)
(13,1.28649493481046)
(14,1.34729157473427)
(15,1.34857960180825)
(16,1.62746797159072)
(17,1.38800056601104)
(18,1.45421496673297)
(19,1.46891687657431)
(20,1.59386542476313)
(21,1.48864451350705)
(22,1.74146544546211)
(23,1.64407939767283)
};
\addplot[black,very thick, loosely dotted] coordinates {
(1,1.1004605753252)
(2,1.11772151594703)
(3,1.12645965981855)
(4,1.13311627925989)
(5,1.14808140394349)
(6,1.16137919758007)
(7,1.16941876825462)
(8,1.19077444425049)
(9,1.19418041088782)
(10,1.23150525152635)
(11,1.22949087415946)
(12,1.25796803721942)
(13,1.28717529942448)
(14,1.29925658607001)
(15,1.32170147004112)
(16,1.36561058857247)
(17,1.38058668009568)
(18,1.42318521552165)
(19,1.47021412712151)
(20,1.47488483099295)
(21,1.59994671876388)
(22,1.5867267859659)
(23,1.62551534585433)
};
\addplot[ForestGreen,very thick,densely dashed] coordinates {
(1,1.09504794320239)
(2,1.09842048786489)
(3,1.10757968883363)
(4,1.10556907413646)
(5,1.12728919477659)
(6,1.13825343173709)
(7,1.14981995428243)
(8,1.16061659172742)
(9,1.18006802677947)
(10,1.18482924446782)
(11,1.19894596177235)
(12,1.25268495634477)
(13,1.24036133053999)
(14,1.26087800630677)
(15,1.28061478188804)
(16,1.35434629552157)
(17,1.33439082838625)
(18,1.34685494223363)
(19,1.39978437211213)
(20,1.51579766536965)
(21,1.43183464259847)
(22,1.54652765810788)
(23,1.5685903500473)
};
\addplot[brown,very thick, loosely dashed] coordinates {
(1,1.09348790877413)
(2,1.09668075931998)
(3,1.11297480092046)
(4,1.11727151923022)
(5,1.12008227148881)
(6,1.14231899191672)
(7,1.14915279001845)
(8,1.15871403485074)
(9,1.1779688166317)
(10,1.186270857977)
(11,1.20548053264078)
(12,1.22300308345362)
(13,1.24910431643006)
(14,1.26082992934906)
(15,1.29382811259555)
(16,1.30739195761338)
(17,1.32674098360656)
(18,1.36896533151403)
(19,1.38451404424559)
(20,1.45623687830305)
(21,1.45814419930328)
(22,1.50928065004381)
(23,1.56209556993529)
};
\addplot[blue,smooth,very thick,loosely dashdotdotted] coordinates {
(1,1.0937470588098)
(2,1.09450618119531)
(3,1.10578786883178)
(4,1.12503865788303)
(5,1.12407064580072)
(6,1.13090958539666)
(7,1.14801745364166)
(8,1.16699405293376)
(9,1.16273767926879)
(10,1.17836718352256)
(11,1.19510999267503)
(12,1.26000357270454)
(13,1.22634852972721)
(14,1.25230904643241)
(15,1.26696452493537)
(16,1.42489965640703)
(17,1.32597720649601)
(18,1.34210429673155)
(19,1.40454944034662)
};
\addplot[orange,smooth,very thick, densely dashdotdotted] coordinates {
(1,1.08846340451563)
(2,1.0983572091778)
(3,1.11157384267296)
(4,1.11416668437945)
(5,1.12355471917102)
(6,1.13740997130219)
(7,1.14315057965412)
(8,1.15747734342568)
(9,1.16852509827638)
(10,1.18575031934045)
(11,1.19589842237852)
(12,1.21745655980481)
(13,1.24563209826368)
(14,1.2504658648145)
(15,1.26200792451755)
(16,1.31469527334052)
(17,1.30546722113503)
(18,1.34535641413809)
(19,1.38850301590203)
};
\end{axis}

\begin{axis}[
at = {(axis1.south east)},
xshift = 0.1\textwidth,
width = 0.5\textwidth, 
height = 0.35\textwidth,
title = {$r=5$},
title style = {align=center, font=\small},
xmin = 1,
xmax = 23,
ymajorgrids,
ymode = log,
xlabel = Value of $i$,
xlabel style = {font=\small},
log ticks with fixed point,
]
\draw (axis cs:\pgfkeysvalueof{/pgfplots/xmin},1.0)  -- (axis cs:\pgfkeysvalueof{/pgfplots/xmax},1.0);
\addplot[red,very thick] coordinates {
(1,1.12564692744292)
(2,1.15507680707956)
(3,1.17070644627032)
(4,1.2007899605943)
(5,1.22813053136229)
(6,1.26275313324656)
(7,1.30259550356425)
(8,1.34579500244732)
(9,1.37899225691668)
(10,1.45750133691188)
(11,1.46107351878879)
(12,1.63913153793178)
(13,1.61166390809074)
(14,1.83515580331618)
(15,1.73361417420394)
(16,0.89118312814001)
(17,1.71915337889142)
(18,1.81404958677686)
(19,2.042194092827)
(20,2.29170024174053)
(21,2.02117263843648)
(22,2.63519313304721)
(23,3.10666666666667)
};
\addplot[black,very thick, loosely dotted] coordinates {
(1,1.13529487840461)
(2,1.15423344948912)
(3,1.17364054715029)
(4,1.19362332279858)
(5,1.21462813061373)
(6,1.24850873807026)
(7,1.27302203744027)
(8,1.2945021163896)
(9,1.33603136541275)
(10,1.36838077945659)
(11,1.41356025421657)
(12,1.45620873155712)
(13,1.4997597693786)
(14,1.56940284439604)
(15,1.60216785798181)
(16,1.70787482806052)
(17,1.74724614460244)
(18,1.76649566601804)
(19,1.92606473594549)
(20,1.99118046132972)
(21,2.02194787379973)
(22,2.06515580736544)
(23,2.824)
};
\addplot[ForestGreen,very thick,densely dashed] coordinates {
(1,1.12720966106489)
(2,1.13761384104825)
(3,1.15453418808965)
(4,1.16968176190725)
(5,1.18936853459948)
(6,1.21266554739592)
(7,1.23788593264091)
(8,1.28349123668981)
(9,1.30592139817336)
(10,1.33116279421324)
(11,1.36662207737877)
(12,1.47659866539562)
(13,1.46084001782531)
(14,1.52200084781687)
(15,1.57426191874433)
(16,1.70173525938039)
(17,1.70466160755769)
(18,1.77244029645896)
(19,1.84595895616924)
(20,2.18066298342541)
(21,2.00887902330744)
(22,2.2984693877551)
(23,2.10752688172043)
};
\addplot[brown,very thick, loosely dashed] coordinates {
(1,1.12681484473568)
(2,1.13910433649386)
(3,1.15439251065153)
(4,1.17590256368265)
(5,1.19788590791564)
(6,1.21594207521291)
(7,1.24337508535483)
(8,1.26618284077412)
(9,1.29273132167281)
(10,1.33527675916469)
(11,1.37250226302858)
(12,1.41939912776611)
(13,1.45213979572254)
(14,1.51925101640831)
(15,1.57713628815941)
(16,1.60529013737954)
(17,1.6983563170358)
(18,1.80218400903728)
(19,1.83740774907749)
(20,1.90845070422535)
(21,1.96539792387543)
(22,2.3640081799591)
(23,2.40886699507389)
};
\addplot[blue,smooth,very thick,loosely dashdotdotted] coordinates {
(1,1.12157852567025)
(2,1.1369099824796)
(3,1.15052751943409)
(4,1.22249948721638)
(5,1.192307424836)
(6,1.21103997731704)
(7,1.2346630896862)
(8,1.27825110319528)
(9,1.2919785116729)
(10,1.32291982189374)
(11,1.36170388281939)
(12,1.48534997702558)
(13,1.43491351838421)
(14,1.47812104083423)
(15,1.54676783157048)
(16,2.04958805355304)
(17,1.70022763088776)
(18,1.83574413371906)
(19,1.80872093023256)
};
\addplot[orange,smooth,very thick, densely dashdotdotted] coordinates {
(1,1.12796318518001)
(2,1.13858198585678)
(3,1.15487229397246)
(4,1.17502220717839)
(5,1.18946845244324)
(6,1.21971534001054)
(7,1.23793381177347)
(8,1.2669054089041)
(9,1.29923679226616)
(10,1.31548221085198)
(11,1.36259905555218)
(12,1.40585666764017)
(13,1.45271011403382)
(14,1.4812918652953)
(15,1.53709681219518)
(16,1.64107025480416)
(17,1.645124355957)
(18,1.72663371773751)
(19,1.78529411764706)
};
\end{axis}
\end{tikzpicture}
\end{subfigure}

\vspace{0.25cm}
\begin{subfigure}{\textwidth}
\caption{Erroneous team identification}
\label{Fig8b}

\begin{tikzpicture}
\begin{axis}[
name = axis1,
width = 0.5\textwidth, 
height = 0.35\textwidth,
title = {$r=3$},
title style = {align=center, font=\small},
xmin = 1,
xmax = 23,
ymajorgrids,
ymode = log,
xlabel = Value of $i$,
xlabel style = {font=\small},
log ticks with fixed point,
]
\draw (axis cs:\pgfkeysvalueof{/pgfplots/xmin},1.0)  -- (axis cs:\pgfkeysvalueof{/pgfplots/xmax},1.0);
\addplot[red,very thick] coordinates {
(1,1.09465668415282)
(2,1.11348729064039)
(3,1.11743283292893)
(4,1.13422406473203)
(5,1.13496960338138)
(6,1.15558041972701)
(7,1.16925306461348)
(8,1.43282592710471)
(9,0.974378618773174)
(10,1.51599904165474)
(11,1.56522738288236)
(12,1.65676891525349)
(13,1.72728060046189)
(14,1.77664159222692)
(15,1.86034022099596)
(16,2.78236223905816)
(17,2.86575606002309)
(18,2.15150982116681)
(19,2.43397640007492)
(20,2.50146670577882)
(21,2.71152869476892)
(22,2.94641313742437)
(23,3.0015243902439)
};
\addplot[black,very thick, loosely dotted] coordinates {
(1,1.10783148389522)
(2,1.11306225921712)
(3,1.12247797033075)
(4,1.13821705497528)
(5,1.1476906325735)
(6,1.1564989682458)
(7,1.17656767253877)
(8,1.39698754156151)
(9,1.21292100196846)
(10,1.4983094026556)
(11,1.54521601069608)
(12,1.61395596648243)
(13,1.67639086369081)
(14,1.71620024014039)
(15,1.79389850101147)
(16,1.9344637209884)
(17,2.00830200105983)
(18,2.0810187767243)
(19,2.1928397521027)
(20,2.28504672897196)
(21,2.49647741400746)
(22,2.66349852907898)
(23,2.73272933182333)
};
\addplot[ForestGreen,very thick,densely dashed] coordinates {
(1,1.09434456675476)
(2,1.10087688512326)
(3,1.10492381000122)
(4,1.11147893239085)
(5,1.12021324452463)
(6,1.13980330491765)
(7,1.13988910343308)
(8,1.3746106459544)
(9,1.37960427324707)
(10,1.44152548539274)
(11,1.46712975908402)
(12,1.54618839125829)
(13,1.60319463942336)
(14,1.61109600628361)
(15,1.67697710074682)
(16,1.76727796908785)
(17,1.87518548746105)
(18,1.93712025234636)
(19,2.02897896202627)
(20,2.18193613900485)
(21,2.19958609271523)
(22,2.34377892370752)
(23,3.0918863578193)
};
\addplot[brown,very thick, loosely dashed] coordinates {
(1,1.09381770208795)
(2,1.0982259753305)
(3,1.11124985219345)
(4,1.11605257601351)
(5,1.12776941687463)
(6,1.13677804967841)
(7,1.14775189872631)
(8,1.35870613687932)
(9,1.39463260425076)
(10,1.44029512711929)
(11,1.47097798099973)
(12,1.51569872456503)
(13,1.56771679116596)
(14,1.6269153497856)
(15,1.68857724851144)
(16,1.76540808088829)
(17,1.83009534733673)
(18,1.92546441403562)
(19,2.04425881146039)
(20,2.10091546656673)
(21,2.25235187324641)
(22,2.37737049764541)
(23,2.46350884326083)
};
\addplot[blue,smooth,very thick,loosely dashdotdotted] coordinates {
(1,1.08782918722434)
(2,1.09809831506278)
(3,1.10505448800437)
(4,1.12993466438551)
(5,1.12183320117254)
(6,1.13132719998922)
(7,1.14214724728652)
(8,1.37008924619218)
(9,1.32499944037229)
(10,1.40611208754079)
(11,1.44457099741305)
(12,1.57309306229019)
(13,1.6156127916224)
(14,1.58006792347191)
(15,1.65149408978919)
(16,1.77504184538132)
(17,1.83481381154245)
(18,1.82969254695538)
(19,2.29237903225806)
};
\addplot[orange,smooth,very thick, densely dashdotdotted] coordinates {
(1,1.09408277914142)
(2,1.10154905039024)
(3,1.10455142011198)
(4,1.11712203077721)
(5,1.1222250592393)
(6,1.13833746224221)
(7,1.14397583622329)
(8,1.3505089525712)
(9,1.37448778683937)
(10,1.41874685552846)
(11,1.4503076555128)
(12,1.50472372163863)
(13,1.56019539333085)
(14,1.5900721545946)
(15,1.63283882462635)
(16,1.717135518591)
(17,1.77779783689747)
(18,1.83425315206174)
(19,1.92673286111814)
};
\end{axis}

\begin{axis}[
at = {(axis1.south east)},
xshift = 0.1\textwidth,
width = 0.5\textwidth, 
height = 0.35\textwidth,
title = {$r=5$},
title style = {align=center, font=\small},
xmin = 1,
xmax = 23,
ymin = 0.9,
ymajorgrids,
ymode = log,
xlabel = Value of $i$,
xlabel style = {font=\small},
log ticks with fixed point,
legend entries={$D(8+6)/S \qquad$,$D(8+6)/R \qquad$,$D(4 \times 7)/S$,$D(4 \times 7)/R \qquad$,$D(4 \times 6)/S \qquad$,$D(4 \times 6)/R$},
legend style = {at={(-0.1,-0.3)},anchor=north,legend columns = 3,font=\small}
]
\draw (axis cs:\pgfkeysvalueof{/pgfplots/xmin},1.0)  -- (axis cs:\pgfkeysvalueof{/pgfplots/xmax},1.0);
\addplot[red,very thick] coordinates {
(1,1.12833045366892)
(2,1.14745005618905)
(3,1.16898733204194)
(4,1.19504059805309)
(5,1.22069806141397)
(6,1.25403970306372)
(7,1.28476037513256)
(8,1.8090841346311)
(9,1.44658862937809)
(10,2.08669735720375)
(11,2.23450062717197)
(12,2.55186397020432)
(13,2.76430297001563)
(14,3.16660236979884)
(15,3.47987340520225)
(16,2.14176109537299)
(17,2.20091423595995)
(18,3.76533333333333)
(19,4.97724810400867)
(20,5.73979591836735)
(21,4.78238341968912)
(22,7.68627450980392)
(23,11.3529411764706)
};
\addplot[black,very thick, loosely dotted] coordinates {
(1,1.13652148198711)
(2,1.15085545333848)
(3,1.17364826089668)
(4,1.19584899141649)
(5,1.21762933095483)
(6,1.24292794150632)
(7,1.2643655530187)
(8,1.69944407437309)
(9,1.50322440068888)
(10,1.92322179190478)
(11,2.05285930564635)
(12,2.18856515278765)
(13,2.33292528747165)
(14,2.49307434608942)
(15,2.67090100230477)
(16,3.00145025621193)
(17,3.22450743173177)
(18,3.51205432937182)
(19,3.94141689373297)
(20,3.96366083445491)
(21,4.63091482649842)
(22,4.61490683229814)
(23,4.73134328358209)
};
\addplot[ForestGreen,very thick,densely dashed] coordinates {
(1,1.12515883571108)
(2,1.13880884778784)
(3,1.15432771639394)
(4,1.16288554509124)
(5,1.18960888472693)
(6,1.21484069152346)
(7,1.2392066843252)
(8,1.67867803958958)
(9,1.69231497793184)
(10,1.83251262265161)
(11,1.92601808138814)
(12,2.16213208013851)
(13,2.33780833134318)
(14,2.36162812309163)
(15,2.58919659025661)
(16,2.76060983654621)
(17,3.00318344608038)
(18,3.36864197530864)
(19,4.0663430420712)
(20,4.18821096173733)
(21,3.92796610169492)
(22,5.06282722513089)
(23,13.4857142857143)
};
\addplot[brown,very thick, loosely dashed] coordinates {
(1,1.12472573373137)
(2,1.14247257119839)
(3,1.15608449296433)
(4,1.17346727965643)
(5,1.1949558818845)
(6,1.21774885117902)
(7,1.24194905412032)
(8,1.65598208409194)
(9,1.74121201998269)
(10,1.83010997147729)
(11,1.92616424032816)
(12,2.05106897252436)
(13,2.20482355027487)
(14,2.37517146776406)
(15,2.58131502288806)
(16,2.79687388987567)
(17,3.03831078518144)
(18,3.24159373560571)
(19,3.58684807256236)
(20,3.96167883211679)
(21,4.34055118110236)
(22,5.24401913875598)
(23,5.52173913043478)
};
\addplot[blue,smooth,very thick,loosely dashdotdotted] coordinates {
(1,1.12048104319463)
(2,1.13293515088567)
(3,1.15533130133277)
(4,1.21938054160001)
(5,1.19117648755178)
(6,1.2140641433387)
(7,1.23613270227506)
(8,1.66038332413539)
(9,1.61857957300891)
(10,1.80613945375163)
(11,1.88812383740213)
(12,2.19969623606797)
(13,2.36605254179461)
(14,2.29277616317126)
(15,2.47477519314392)
(16,3.00472032098183)
(17,3.2057111923129)
(18,3.07623426911907)
(19,5.10995850622407)
};
\addplot[orange,smooth,very thick, densely dashdotdotted] coordinates {
(1,1.12592436060939)
(2,1.1416277345643)
(3,1.15493608350897)
(4,1.17287537982304)
(5,1.18928378647121)
(6,1.21764837743744)
(7,1.23986528797868)
(8,1.63270662580743)
(9,1.69689858754951)
(10,1.81796793840262)
(11,1.91079346266163)
(12,2.00377547167886)
(13,2.13381172204702)
(14,2.32853375151472)
(15,2.49760273972603)
(16,2.61409937888199)
(17,2.93074606891937)
(18,3.15624387375025)
(19,3.40562096467907)
};
\end{axis}
\end{tikzpicture}
\end{subfigure}

\end{figure}


%% file: Figures_Appendix.tex
\begin{figure}[t]
\centering
\caption{Design $D(8+6)$, which has been used in the men's \\
handball \href{https://en.wikipedia.org/wiki/EHF_Champions_League}{EHF Champions League} since the \href{https://en.wikipedia.org/wiki/2015\%E2\%80\%9316_EHF_Champions_League}{2015/16 season}}
\label{Fig_A1}

\begin{subfigure}{\textwidth}
  \centering
  \subcaption{Group stage}
  \label{Fig_A1a}

\begin{tikzpicture}[scale=1, auto=center, transform shape, >=triangle 45]
  \path
    (-6,0) coordinate (A) node {
    \begin{tabular}{c} \toprule
    \textbf{Group A} \\ \midrule
    A1 \\ \midrule
    A2 \\
    A3 \\ 
    A4 \\ 
    A5 \\
    A6 \\ \midrule
    A7 \\ 
    A8 \\ \bottomrule
    \end{tabular}
    }
    (-2,0) coordinate (B) node { 
    \begin{tabular}{c} \toprule
    \textbf{Group B} \\ \midrule
    B1 \\ \midrule
    B2 \\
    B3 \\ 
    B4 \\
    B5 \\
    B6 \\ \midrule
    B7 \\ 
    B8 \\ \bottomrule
    \end{tabular}
    }
    (2,0) coordinate (A) node {
    \begin{tabular}{c} \toprule
    \textbf{Group C} \\ \midrule
    C1 \\
    C2 \\ \midrule
    C3 \\ 
    C4 \\
    C5 \\
    C6 \\ \bottomrule
    \end{tabular}
    }
    (6,0) coordinate (D) node {
    \begin{tabular}{c} \toprule
    \textbf{Group D} \\ \midrule
    D1 \\
    D2 \\ \midrule
    D3 \\ 
    D4 \\
    D5 \\
    D6 \\ \bottomrule
    \end{tabular}
    };
\end{tikzpicture}
\end{subfigure}

\vspace{0.5cm}
\begin{subfigure}{\textwidth}
  \centering
  \subcaption{Play-off}
  \label{Fig_A1b}
  
  \begin{tikzpicture}[
  level distance=4cm,every node/.style={minimum width=2cm,inner sep=0pt},
  edge from parent/.style={ultra thick,draw},
  level 1/.style={sibling distance=6cm},
  level 2/.style={sibling distance=3cm},
  level 3/.style={sibling distance=1.5cm},
  legend/.style={inner sep=3pt}
]
\node (1) {\Pair{K1}{C2}{D1}};
\node[legend] at ([xshift=8cm]1) {{\Pair{K2}{C1}{D2}}};
\end{tikzpicture}
\end{subfigure}

\vspace{0.5cm}
\begin{subfigure}{\textwidth}
  \centering
  \subcaption{Knockout stage}
  \label{Fig_A1c}
  
  \begin{tikzpicture}[
  level distance=6cm,every node/.style={minimum width=2cm,inner sep=0pt},
  edge from parent/.style={ultra thick,draw},
  level 1/.style={sibling distance=3cm},
  level 2/.style={sibling distance=1.5cm},
  legend/.style={inner sep=3pt}
]
\node (1) {
\begin{tabular}{|>{\centering\arraybackslash}m{0.75cm}|>{\centering\arraybackslash}m{1.5cm}|}
  \hline
  \multirow{4}{*}{F4} & $\mathcal{W}$/QF1 \\ \cline{2-2}
   & $\mathcal{W}$/QF2 \\ \cline{2-2}
   & $\mathcal{W}$/QF3 \\ \cline{2-2}
   & $\mathcal{W}$/QF4 \\ \hline
  \end{tabular}
}
[edge from parent fork left,grow=left]
child {node (2) {\Pair{QF1}{$\mathcal{W}$/M1}{$\mathcal{W}$/M4}}
child {node (3) {\Pair{M1}{$\mathcal{W}$/K1}{A2}}
} 
child {node {\Pair{M4}{A6}{B3}}
} 
} 
child {node {\Pair{QF2}{$\mathcal{W}$/M2}{$\mathcal{W}$/M3}}
child {node {\Pair{M2}{$\mathcal{W}$/K2}{B2}}
} 
child {node {\Pair{M3}{A3}{B6}}
} 
}
child {node {\Pair{QF3}{$\mathcal{W}$/M6}{A1}}
child {node {\Pair{M6}{A5}{B4}}
} 
child {node {A1}
} 
} 
child {node {\Pair{QF4}{$\mathcal{W}$/M5}{B1}}
child {node {\Pair{M5}{A4}{B5}}
} 
child {node {B1}
} 
};
\node[legend] at ([yshift=1cm]3) (QF) {\textbf{Phase 1}};
\node[legend] at (2|-QF) {\textbf{Quarter-finals}};
\node[legend] at (1|-QF) {\textbf{Final Four}};
\end{tikzpicture}
\end{subfigure}

\end{figure}

\begin{figure}[ht!]
\centering
\caption{Design $D(4 \times 7)$, a traditional alternative format with $28$ teams}
\label{Fig_A2}

\begin{subfigure}{\textwidth}
  \centering
  \subcaption{Group stage}
  \label{Fig_A2a}

\begin{tikzpicture}[scale=1, auto=center, transform shape, >=triangle 45]
  \path
    (-6,0) coordinate (A) node {
    \begin{tabular}{c} \toprule
    \textbf{Group A} \\ \midrule
    A1 \\
    A2 \\ \midrule
    A3 \\ 
    A4 \\ 
    A5 \\
    A6 \\ \midrule
    A7 \\ \bottomrule
    \end{tabular}
    }
    (-2,0) coordinate (B) node { 
    \begin{tabular}{c} \toprule
    \textbf{Group B} \\ \midrule
    B1 \\
    B2 \\ \midrule
    B3 \\ 
    B4 \\
    B5 \\
    B6 \\ \midrule
    B7 \\ \bottomrule
    \end{tabular}
    }
    (2,0) coordinate (A) node {
    \begin{tabular}{c} \toprule
    \textbf{Group C} \\ \midrule
    C1 \\
    C2 \\ \midrule
    C3 \\ 
    C4 \\
    C5 \\
    C6 \\ \midrule
    C7 \\ \bottomrule
    \end{tabular}
    }
    (6,0) coordinate (D) node {
    \begin{tabular}{c} \toprule
    \textbf{Group D} \\ \midrule
    D1 \\
    D2 \\ \midrule
    D3 \\ 
    D4 \\
    D5 \\
    D6 \\ \midrule
    D7 \\ \bottomrule
    \end{tabular}
    };
\end{tikzpicture}
\end{subfigure}

\vspace{0.5cm}
\begin{subfigure}{\textwidth}
  \centering
  \subcaption{Knockout stage}
  \label{Fig_A2b}
  
  \begin{tikzpicture}[
  level distance=4cm,every node/.style={minimum width=2cm,inner sep=0pt},
  edge from parent/.style={ultra thick,draw},
  level 1/.style={sibling distance=4cm},
  level 2/.style={sibling distance=2cm},
  level 3/.style={sibling distance=1cm},
  legend/.style={inner sep=3pt}
]
\node (1) {
\begin{tabular}{|>{\centering\arraybackslash}m{0.75cm}|>{\centering\arraybackslash}m{1.5cm}|}
  \hline
  \multirow{4}{*}{F4} & $\mathcal{W}$/QF1 \\ \cline{2-2}
   & $\mathcal{W}$/QF2 \\ \cline{2-2}
   & $\mathcal{W}$/QF3 \\ \cline{2-2}
   & $\mathcal{W}$/QF4 \\ \hline
  \end{tabular}
}
[edge from parent fork left,grow=left]
child {node (2) {\Pair{QF1}{$\mathcal{W}$/L1}{$\mathcal{W}$/L8}}
child {node (3) {\Pair{L1}{$\mathcal{W}$/K6}{A1}}
child {node (4) {\Pair{K6}{C4}{D5}}}
child {node {A1}}
} 
child {node {\Pair{L8}{$\mathcal{W}$/K4}{D2}}
child {node {\Pair{K4}{A6}{B3}}}
child {node {D2}
} 
} 
} 
child {node {\Pair{QF2}{$\mathcal{W}$/L3}{$\mathcal{W}$/L6}}
child {node {\Pair{L3}{$\mathcal{W}$/K7}{B1}}
child {node {\Pair{K7}{C5}{D4}}}
child {node {B1}}
} 
child {node {\Pair{L6}{$\mathcal{W}$/K1}{C2}}
child {node {\Pair{K1}{A3}{B6}}}
child {node {C2}
} 
} 
}
child {node {\Pair{QF3}{$\mathcal{W}$/L2}{$\mathcal{W}$/L7}}
child {node {\Pair{L2}{$\mathcal{W}$/K5}{A2}}
child {node {\Pair{K5}{C3}{D6}}}
child {node {A2}}
} 
child {node {\Pair{L7}{$\mathcal{W}$/K3}{D1}}
child {node {\Pair{K3}{A5}{B4}}}
child {node {D1}
} 
} 
} 
child {node {\Pair{QF4}{$\mathcal{W}$/L4}{$\mathcal{W}$/L5}}
child {node {\Pair{L4}{$\mathcal{W}$/K8}{B2}}
child {node {\Pair{K8}{C6}{D3}}}
child {node {B2}}
} 
child {node {\Pair{L5}{$\mathcal{W}$/K2}{C1}}
child {node {\Pair{K2}{A4}{B3}}}
child {node {C1}
} 
} 
};
\node[legend] at ([yshift=1cm]4) (R16) {\textbf{Phase 1}};
\node[legend] at (3|-R16) {\textbf{Round of 16}};
\node[legend] at (2|-R16) {\textbf{Quarter-finals}};
\node[legend] at (1|-R16) {\textbf{Final Four}};
\end{tikzpicture}
\end{subfigure}
\end{figure}

\begin{figure}[ht!]
\centering
\caption{Design $D(4 \times 6)$, the previous format \\
of the men's handball \href{https://en.wikipedia.org/wiki/EHF_Champions_League}{EHF Champions League}}
\label{Fig_A3}

\begin{subfigure}{\textwidth}
  \centering
  \subcaption{Group stage}
  \label{Fig_A3a}

\begin{tikzpicture}[scale=1, auto=center, transform shape, >=triangle 45]
  \path
    (-6,0) coordinate (A) node {
    \begin{tabular}{c} \toprule
    \textbf{Group A} \\ \midrule
    A1 \\
    A2 \\
    A3 \\ 
    A4 \\ \midrule
    A5 \\
    A6 \\ \bottomrule
    \end{tabular}
    }
    (-2,0) coordinate (B) node { 
    \begin{tabular}{c} \toprule
    \textbf{Group B} \\ \midrule
    B1 \\
    B2 \\
    B3 \\ 
    B4 \\ \midrule
    B5 \\
    B6 \\ \bottomrule
    \end{tabular}
    }
    (2,0) coordinate (A) node {
    \begin{tabular}{c} \toprule
    \textbf{Group C} \\ \midrule
    C1 \\
    C2 \\
    C3 \\ 
    C4 \\ \midrule
    C5 \\
    C6 \\ \bottomrule
    \end{tabular}
    }
    (6,0) coordinate (D) node {
    \begin{tabular}{c} \toprule
    \textbf{Group D} \\ \midrule
    D1 \\
    D2 \\
    D3 \\ 
    D4 \\ \midrule
    D5 \\
    D6 \\ \bottomrule
    \end{tabular}
    };
\end{tikzpicture}
\end{subfigure}

\vspace{0.5cm}
\begin{subfigure}{\textwidth}
  \centering
  \subcaption{Knockout stage}
  \label{Fig_A3b}
  
  \begin{tikzpicture}[
  level distance=6cm,every node/.style={minimum width=2cm,inner sep=0pt},
  edge from parent/.style={ultra thick,draw},
  level 1/.style={sibling distance=6cm},
  level 2/.style={sibling distance=1.5cm},
  legend/.style={inner sep=5pt}
]
\node (1) {
\begin{tabular}{|>{\centering\arraybackslash}m{0.75cm}|>{\centering\arraybackslash}m{1.5cm}|}
  \hline
  \multirow{4}{*}{F4} & $\mathcal{W}$/QF1 \\ \cline{2-2}
   & $\mathcal{W}$/QF2 \\ \cline{2-2}
   & $\mathcal{W}$/QF3 \\ \cline{2-2}
   & $\mathcal{W}$/QF4 \\ \hline
  \end{tabular}
}
[edge from parent fork left,grow=left]
child {node (2) {
\begin{tabular}{|>{\centering\arraybackslash}m{1cm}|>{\centering\arraybackslash}m{1.5cm}|}
  \hline
  \multirow{2}{*}{QF} & $\mathcal{W}$/K1 \\ \cline{2-2}
   & $\mathcal{W}$/K2 \\ \cline{2-2}
  \multirow{2}{*}{Pot 1} & $\mathcal{W}$/K3 \\ \cline{2-2}
   & $\mathcal{W}$/K4 \\ \hline
  \end{tabular}
}
child {node (3) {\WidePair{K1}{A1}{B4/C4/D4}}
} 
child {node {\WidePair{K2}{B1}{A4/C4/D4}}
} 
child {node {\WidePair{K3}{C1}{A4/B4/D4}}
} 
child {node {\WidePair{K4}{D1}{A4/B4/C4}}
} 
}
child {node (2) {
\begin{tabular}{|>{\centering\arraybackslash}m{1cm}|>{\centering\arraybackslash}m{1.5cm}|}
  \hline
  \multirow{2}{*}{QF} & $\mathcal{W}$/K5 \\ \cline{2-2}
   & $\mathcal{W}$/K6 \\ \cline{2-2}
  \multirow{2}{*}{Pot 2} & $\mathcal{W}$/K7 \\ \cline{2-2}
   & $\mathcal{W}$/K8 \\ \hline
  \end{tabular}
}
child {node (3) {\WidePair{K5}{A2}{B3/C3/D3}}
} 
child {node {\WidePair{K6}{B2}{A3/C3/D3}}
} 
child {node {\WidePair{K7}{C2}{A3/B3/D3}}
} 
child {node {\WidePair{K8}{D2}{A3/B3/C3}}
} 
};
\node[legend] at ([yshift=-2cm]3|-R16) {\textbf{Round of 16}};
\node[legend] at ([yshift=-2cm]2|-R16) {\textbf{Quarter-finals}};
\node[legend] at ([yshift=-2cm]1|-R16) {\textbf{Final Four}};
\end{tikzpicture}
\end{subfigure}
\end{figure}
